%% file: paper-arxiv.tex
\documentclass{article}

\usepackage{a4wide, graphicx, siunitx}
\usepackage{float}
\usepackage{subcaption}
\usepackage{lscape}
\usepackage{longtable}
\usepackage{xcolor}
\usepackage{colortbl}
\usepackage{multirow}
\usepackage{gensymb}
\usepackage{pbox}
\usepackage{float}
\usepackage[bottom]{footmisc}
\usepackage{amsmath}
\usepackage[export]{adjustbox}
\begin{document}
\title{\vspace{-3ex} \uppercase{\rmfamily \large Benchmarking
    high-fidelity pedestrian tracking systems\\ for
    research, real-time monitoring and crowd control}}
\author{\normalsize\scshape Caspar A.S. Pouw$^{\,a,b}$, Joris
  Willems$^{\,c}$, Frank van Schadewijk$^{\,b}$,\\\normalsize\scshape Jasmin Thurau$^{\,d}$, Federico
  Toschi$^{\,a,e}$ and Alessandro Corbetta$^{\,a}$\\[7pt]
  \footnotesize \parbox[t]{0.96\linewidth}{\centering
    $^{\,a}$Department of Applied Physics, Eindhoven University of
    Technology, 5600 MB Eindhoven, The Netherlands}\\
  \footnotesize \parbox[t]{0.96\linewidth}{\centering
    $^{\,b}$ProRail Stations, Utrecht, The Netherlands}\\
   \footnotesize \parbox[t]{0.96\linewidth}{\centering
    $^{\,c}$Department of Mathematics
    and Computer Science, Eindhoven University of Technology 5600 MB
    Eindhoven, The Netherlands}\\
   \footnotesize \parbox[t]{0.96\linewidth}{\centering
     $^{\,d}$SBB AG, CH-3000 Bern, Switzerland}\\
   \footnotesize \parbox[t]{0.96\linewidth}{\centering
    $^{\,e}$CNR-IAC, I-00185 Rome, Italy}}
 
%\date{\vspace{1ex}} \setlength\textheight{700pt}

% \begin{frontmatter}
% \title{Benchmarking the state-of-the-art in high-fidelity pedestrian tracking for research, real-time monitoring and crowd control}
% % \author{Caspar A.S. Pouw, Joris Willems, Frank van Schadewijk,\\ Jasmin Thurau,
% %   Federico Toschi, Alessandro Corbetta}

% \author[AP,ProRail]{Caspar A.S. Pouw}

% \author[AP,CNR-IAC]{Federico Toschi}
% \author[ProRail]{Frank van Schadewijk}

% \author[AP]{Alessandro Corbetta}
% %\author[AP]{Alessandro Corbetta \corref{mycorrespondingauthor}}
% %\cortext[mycorrespondingauthor]{Corresponding author}
% %\ead{a.corbetta@tue.nl}

% \address[AP]{Department of Applied Physics, Eindhoven University of Technology 5600 MB Eindhoven, The Netherlands}

% \address[ProRail]{ProRail Stations, 3511 EP Utrecht, The Netherlands}
% \address[CNR-IAC]{CNR-IAC I-00185, Rome, Italy}
% %\date{May 2021}

\maketitle

\begin{abstract}
  High-fidelity pedestrian tracking in real-life conditions has been
  an important tool in fundamental crowd dynamics research allowing to
  quantify statistics of relevant observables including walking
  velocities, mutual distances and body orientations. As this
  technology advances, it is becoming increasingly useful also in
  society. In fact, continued urbanization is overwhelming existing
  pedestrian infrastructures such as transportation hubs and stations,
  generating an urgent need for real-time highly-accurate usage data,
  aiming both at flow monitoring and dynamics understanding. To
  successfully employ pedestrian tracking techniques in research and
  technology, it is crucial to validate and benchmark them for
  accuracy. This is not only necessary to guarantee data quality, but
  also to identify systematic errors. Currently, there is no
  established policy in this context.

  In this contribution, we present and discuss a benchmark suite,
  towards an open standard in the community, for privacy-respectful
  pedestrian tracking techniques. The suite is technology-independent
  and it is applicable to academic and commercial pedestrian tracking
  systems, operating both in lab environments and real-life
  conditions. The benchmark suite consists of 5 tests addressing
  specific aspects of pedestrian tracking quality, including accurate
  line-based crowd flux estimation, local density estimation,
  individual position detection and trajectory accuracy. The output of the tests are
  quality factors expressed as single numbers. We provide the
  benchmark results for two tracking systems, both operating in
  real-life, one commercial, and the other based on overhead
  depth-maps developed at TU Eindhoven, within the Crowdflow topical
  group. We discuss the results on the basis of the quality factors
  and report on the typical sensor and algorithmic performance. This
  enables us to highlight the current state-of-the-art, its
  limitations and provide installation recommendations, with specific
  attention to multi-sensor setups and data stitching.
  \\[0.5cm]
\end{abstract}

\section{Introduction}
\label{sec:intro}
Growing population and continued urbanization puts urban
infrastructures at large stress. Moreover, over the next 10 years in
densely populated European countries public transport facilities such
as, e.g. train or metro stations, expect a passenger growth as high as
40\%~\cite{piereringaprorail}. Potentially dangerous crowd-capacity
issues -- possibly in combination with distancing requirements --
increase by the day, and demand substantial crowd management
efforts. To unlock sustainable and scalable crowd management,
maximizing comfort and safety, real-time, high-accuracy, anonymous
individual pedestrian tracking is a must. This enables reliable usage
monitoring and performance profiling, and, on a broader perspective,
the possibility to develop a fundamental understanding of the motion
of crowd flows. Pedestrian dynamics researchers, who historically
mostly relied on controlled laboratory experiments (see
e.g.~\cite{Daamen2003, Seyfried2005, Kretz2006, Moussad2009,
  Schadschneider2011, Seitz2012, Yamamoto2019}), can now also acquire
fundamental knowledge in real-life environments collecting large-scale
statistics of observables such as walking velocities, mutual
distances, body orientations and group structure~\cite{Corbetta2014,
  Brscic2014, Zanlungo2017, Corbetta2018, Willems2020, Pouw2020}.

Optic-based tracking, leveraging on visual-like signals, is the most
widespread technology when (sub-)centimeter pedestrian positioning
resolution is required. Raw signals are generally acquired via arrays
of color cameras (as CCTV)~\cite{Daamen2003, Moussad2009}, stereo
cameras~\cite{Seyfried2005, Pouw2020, Boltes2013} or infrared
depth-cameras~\cite{Corbetta2014, Brscic2013, Seer2014}. Specifically,
the last two technologies, hinged on three-dimensional imaging, allow
higher accuracy and will be considered in this paper. After a
calibration stage in which, among others, pixel-coordinates are
matched to spatial coordinates, raw signals are post-processed to
yield pedestrian positions and trajectories.  Highly accurate, optical
methods are generally limited in range by the visual cone of the
individual sensors. Note that, at the price of a substantial accuracy loss,
Bluetooth-based~\cite{Yoshimura2014, Centorrino2021} or
Wi-Fi-based~\cite{Hong2018, Georgievska2019} tracking enable larger
spatial coverage per sensor.

Optic-based tracking techniques become rapidly more affordable over
time, this makes high-accuracy pedestrian tracking accessible to a
wide variety of users beyond academic research. Over the last few
years, for instance, managers of public transport facilities have
adopted pedestrian tracking technologies to gain valuable insights in
the flow dynamics through their facilities (see e.g.~\cite{ Pouw2020, Heuvel2019,
  Thurau2019, Thurau2020}).

To successfully employ pedestrian tracking techniques in research and
technology, it is crucial to measure the level of performance and
asses the quality of the output data. This means establishing a
benchmark as well as measuring the performance of the current
state-of-the-art. This is not only necessary to guarantee data
quality, but also to identify systematic errors, such as erroneous
object-person recognition, and misaligned stitching. To the best of our
knowledge, there is no established tool in this context.

In this contribution we propose a benchmark that consists of a minimal
set of five tests to quantify pedestrian detection accuracy and
time-tracking reliability. The tests have been iteratively improved
over the past 3 years~\cite{Heuvel2019} and follow from a joint effort
between academic research and large-scale public facility
managers. The full benchmark, designed to take limited time and
resources, can be executed in less than $2$ hours with as few as $12$
participants.  The five distinct tests span from macroscopic to
increasingly microscopic observables of pedestrian
dynamics. Considering finer and finer aspects of pedestrian dynamics,
the tests get increasingly challenging from a technological
perspective. The tests output quality factors expressed as single
numbers. The first two tests probe large-scale observables by gauging
reliability in estimating: 1. crowd fluxes (ped/min), and 2. local
densities (ped/m$^2$). Note that these are averaged quantities,
respectively over a time interval or over a surface, and therefore
benefit from error compensation, i.e. false negatives could
counterbalance false positives. In test 3, we focus on the
significantly more challenging task of instantaneous individual
localization. In tests 4 and 5, we consider Lagrangian time-tracking
proficiency over full multi-sensor measurement domains.

We present the benchmark results for two pedestrian tracking setups
operating in real-life, one developed in-house at TU Eindhoven, and
the other one commercially available. This aims at reporting on the
current state-of-the-art and providing a reference for new pedestrian
tracking setups.

This paper is structured as follows: In Section~\ref{sec: essentials}
we discuss the essentials of optic-based pedestrian tracking. This is
followed, in Section~\ref{sec: setups}, by a description of the two
experimental setups we employ to validate our benchmark. In
Section~\ref{sec: benchmark}, we introduce the benchmark suite and
detail individual tests and their rationale. In Section~\ref{sec:
  results}, we report the benchmark performance of our experimental
setups. The discussion in Section~\ref{sec: discussion} concludes the
paper.

\section{Essentials of 3D optic-based pedestrian tracking}\label{sec:
  essentials}
In 3D optic-based tracking, visible-light or infrared 3D imaging data,
acquired by camera-like sensors, are processed to localize pedestrians
in space and track them over time. Three-dimensional imaging, richer
in information than flat 2D pictures, substantially simplifies and
allows high accuracy in the tasks connected to tracking. At the core
it is the estimation of a depth-map of the scene, that encodes the
position of each pixel in the three-dimensional
space~\cite{Corbetta2018,Brscic2013,Seer2014}. This can be achieved
via stereoscopic vision, scattered infrared illumination
e.g.~\cite{kinect}, or time-of-flight sensors. For each pedestrian and
each 3D-frame acquired, the tracking process yields quadruplets
$(x,y,t,id)$ where $x$ and $y$ are spatial coordinates (a
$z$-coordinate can be added if needed), $t$ is the frame acquisition
time, and $id$ is an identifier unique to each pedestrian. This
enables us to define the set of recorded trajectories
$\mathcal{T} = \left\{ \gamma_{id} \right\}$, with $\gamma_{id}$ the
trajectory of the pedestrian with identifier $id$.

The tracking process generally happens in two stages: localization and
Lagrangian time-tracking. In the localization stage, each frame is
processed independently to single out each pedestrian and estimate
their position. To this purpose image processing and/or machine
learning models are used~\cite{Willems2020, Brscic2013, Seer2014, Kroneman2020}. Lagrangian time-tracking assigns an $id$ to each
detection on the basis of continuity arguments. These two stages can
also happen simultaneously, e.g. using optical-flow like
techniques~\cite{Boltes2013}.

Cameras or infrared sensors are generally mounted in overhead
position, aimed perpendicularly to the floor to reduce mutual
pedestrians occlusions. The scope of their view-cone depends on, and
in general, limited to, the mounting height (typically the height of
the ceiling). Grids of sensors with partially overlapping view-cones
can be combined to enlarge the measurement domain~\cite{Corbetta2020}.
During the installation, a calibration step which establishes a global
coordinate system across all the sensors is generally performed.
Additionally, background subtraction which removes stationary objects
(e.g. benches) from the image can be used to simplify the localization
phase and increase its accuracy.

Despite its growing adoption, optic-based pedestrian tracking still
features numerous open technical challenges. First, localization
algorithms can fail due to poor image quality, e.g. caused by
impurities on the camera lenses, excessive or insufficient
illumination, or interference in the infrared spectrum (caused by
direct sun exposure).  Additionally, objects in the scene such as
luggage or bicycles can yield false positive detections. Instead,
possible causes for false negatives are (partial) occlusions by other
pedestrians or infrastructural objects, e.g. trusses, signage, and
lighting. Especially at the boundaries of each sensors view-cone the
acquired image can be distorted which also lowers the localization
quality. False negatives in the localization process yield gaps over
which the time-tracking algorithm is unable to return continuous
trajectories. As a consequence, two or more ``broken'' trajectory
pieces with distinct $id$s are returned.

Combining information from multiple sensors presents also challenges
connected to insufficient overlap or misalignment between the sensor
view-cones (cf., e.g., the view-cone stitching algorithm in
~\cite{Corbetta2018}). These are typical causes for ``broken'' trajectories.

Our benchmark contains tests that are designed to recognize these
bottlenecks, probing efficiently localization accuracy, geometric
conformality, and the absence of tracking artifacts such as
misassigned $id$s.

\section{Tracking technologies and experimental setups}\label{sec: setups}
We discuss our benchmark considering two pedestrian tracking setups
briefly described below. The systems, which operate anonymously and in
real-life conditions, leverage on different optic-based tracking
approaches. The first, developed in house at TU Eindhoven, within the Crowdflow
topical group, is based on depth reconstructed via scattered infrared
illumination~\cite{kinect} and, the second, is based on depth reconstructed via
stereoscopic vision. In the following we shall refer to these
systems, respectively, as ``TU/e setup'' and ``commercial setup''. We
give a description of the pedestrian tracking systems in the
paragraphs below. Note that in both setups we did not apply any
further processing of the raw images nor post-processed the obtained
trajectories. The performance of these tracking systems can be
enhanced by setup-specific manual operations e.g. smoothing or
restitching of the individual trajectories. This choice was
deliberately made to focus on the bare, baseline, tracking accuracy.
 
\begin{figure}[h]
  \centering
    \includegraphics[width=0.4\linewidth]{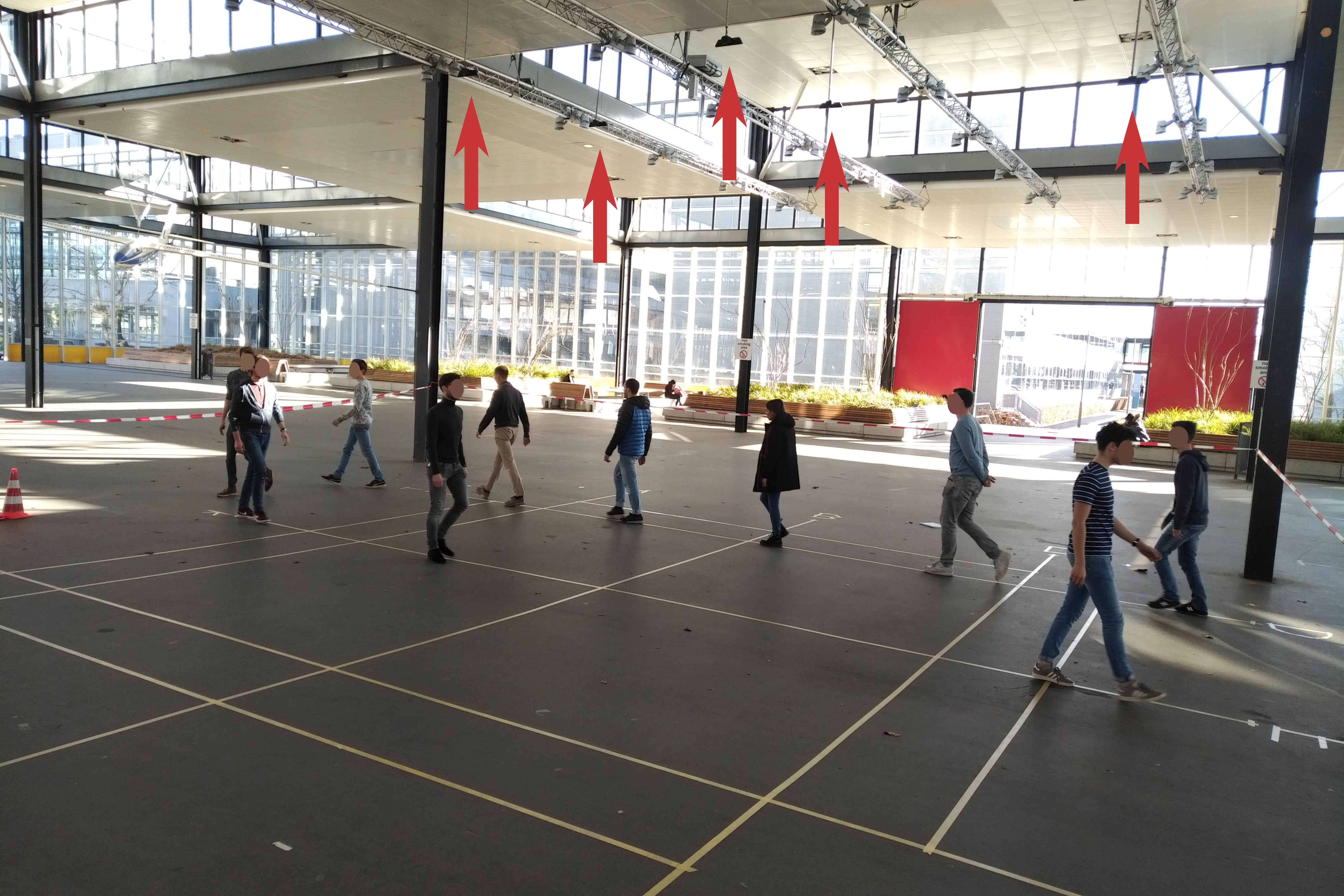}
    \includegraphics[trim = 0 0 800 500, clip,
    width=0.33\linewidth]{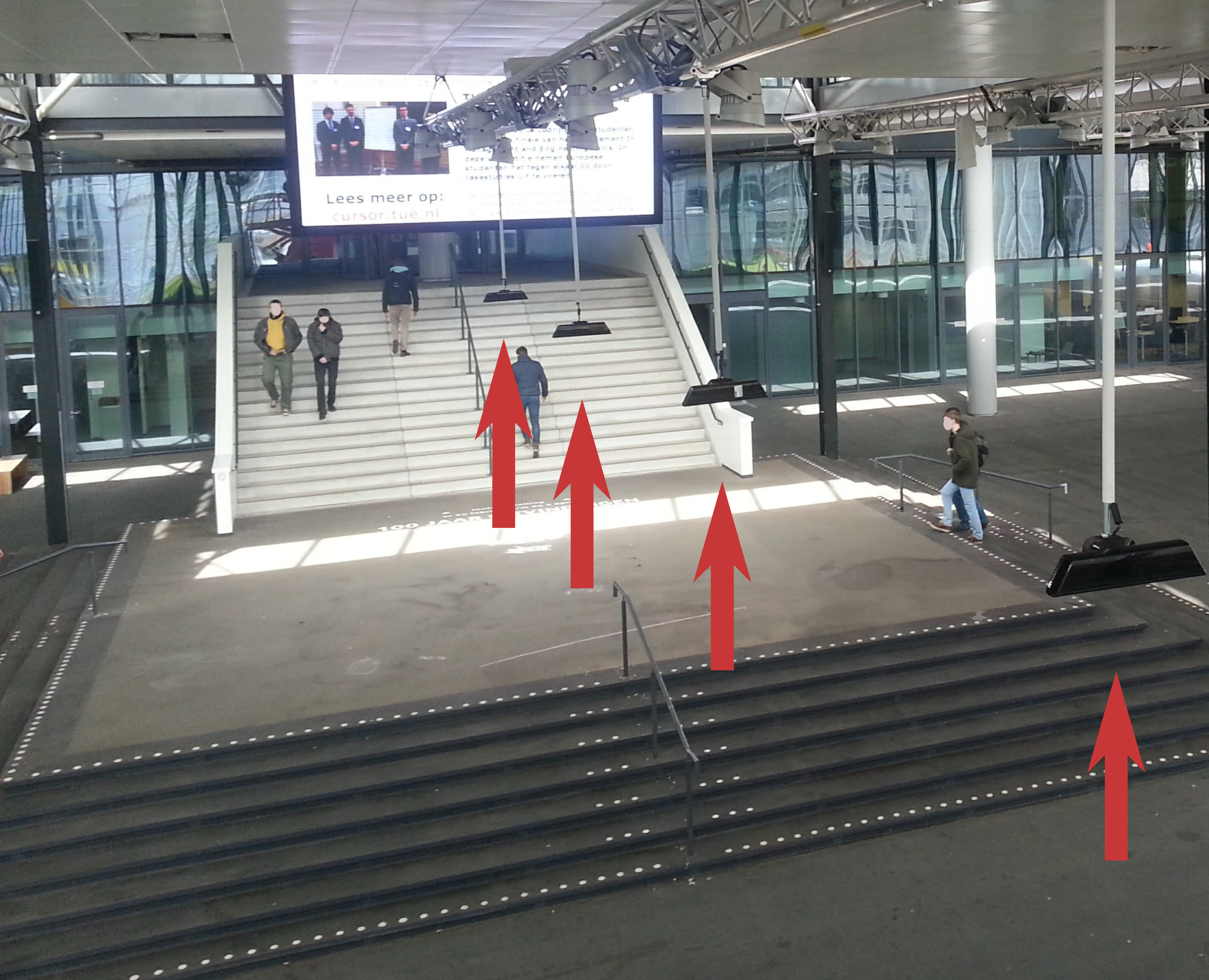}
    \includegraphics[trim = 0 0 0 0, clip, frame,
    width=0.21\linewidth]{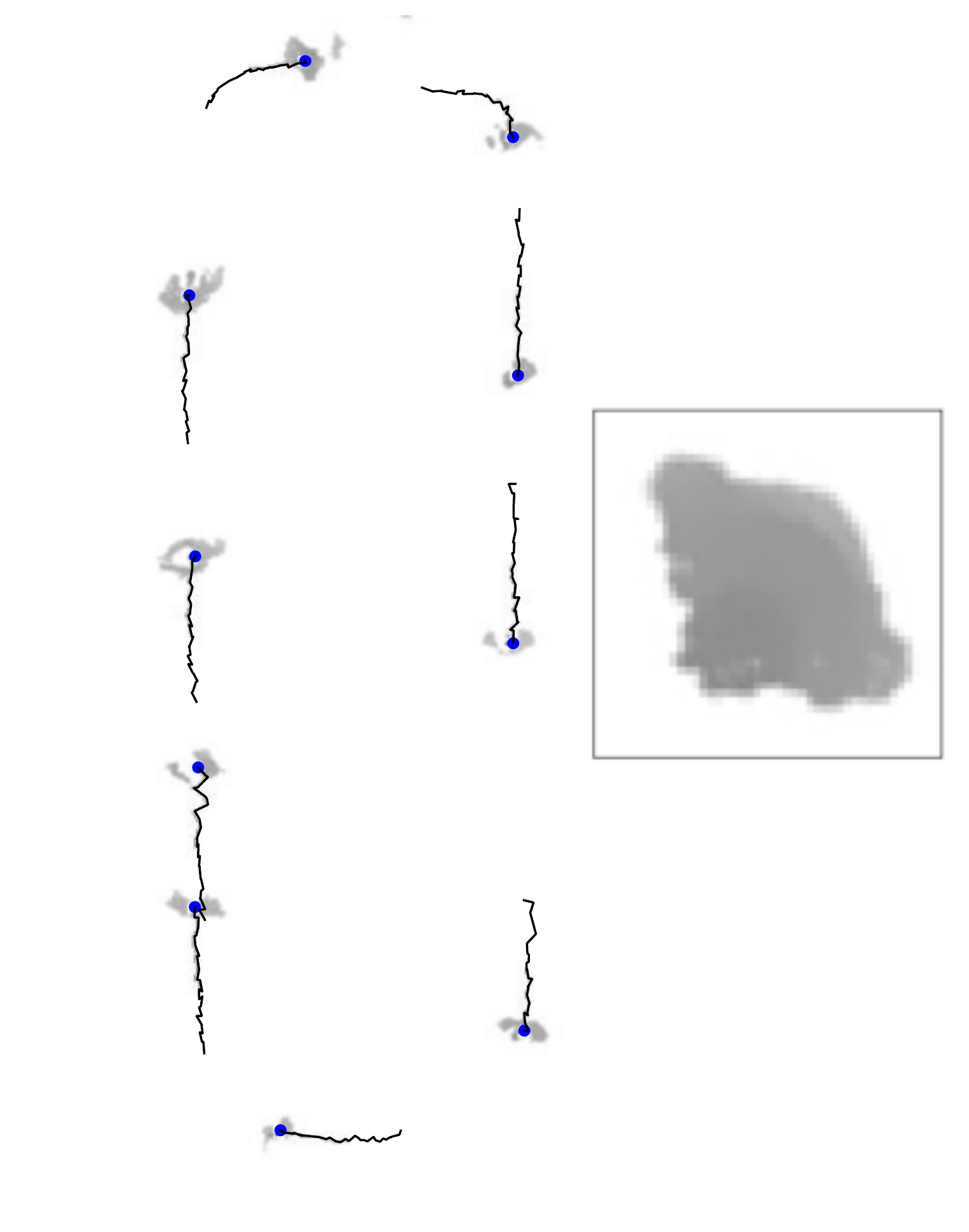}
  \caption{TU/e measurement setup based on overhead depth sensors at
    Eindhoven University of Technology, The Netherlands, during test
    3. (a) Picture taken during test 3 of the benchmark. At the top of
    the image part of the $3\times 4$ sensor grid is indicated with
    red arrows. We guide participants during this test using taped
    markings on the floor. (b) Detail of one of the three arrays of 4
    sensors, taken at sensor height. The sensors are attached to a
    truss near the ceiling pointing downward, perpendicularly to the
    floor (c) Overhead depth image captured by the sensor grid already
    encompassing merging and perspective
    correction~\cite{Corbetta2017}. The gray color in the depth-map
    represents the distance to the camera plane. Bright shades are far
    from the sensor and darker colors are closer to the sensor. In the
    depth-map we can distinguish silhouettes of pedestrians (as seen
    from above) where the shoulders are a lighter tint of gray and the
    head is slightly darker. The inset reports the silhouette of a
    single pedestrian. The trajectories of the pedestrians are
    super-imposed and show the walking direction.}
  \label{fig: setup1}
\end{figure}

\paragraph{TU/e setup}
The tracking system developed in house at TU Eindhoven leverages on
overhead depth-map images, which represent the distance between pixels
and the sensor plane (colorized in shades of gray in the example of
Fig.~\ref{fig: setup1}c). Localization occurs via depth clustering (as
in~\cite{Corbetta2018}), and time-tracking uses the Trackpy Python
library~\cite{trackpy}. The same approach has been successfully used
in e.g. in stations, streets, and museums~\cite{Corbetta2014,
  Corbetta2018, Willems2020, Kroneman2020}. The specific setup
considered consists of a grid of $3 \times 4$ Microsoft
Kinect\textsuperscript{TM}~\cite{kinect} depth sensors. The sensors
are attached to the ceiling of a large public area within the
University campus in Eindhoven, the Netherlands, at a height of about
$4.5\;$m (see Fig.~\ref{fig: setup1}a, b). The grid records depth
images (Fig.~\ref{fig: setup1}c) over an area of
$S = 150\;\textrm{m}^2$ with $f = 30$ frames per second.

\begin{figure}[h]
  \centering
  \includegraphics[trim = 0 60 0 50, clip, width=0.9\linewidth]{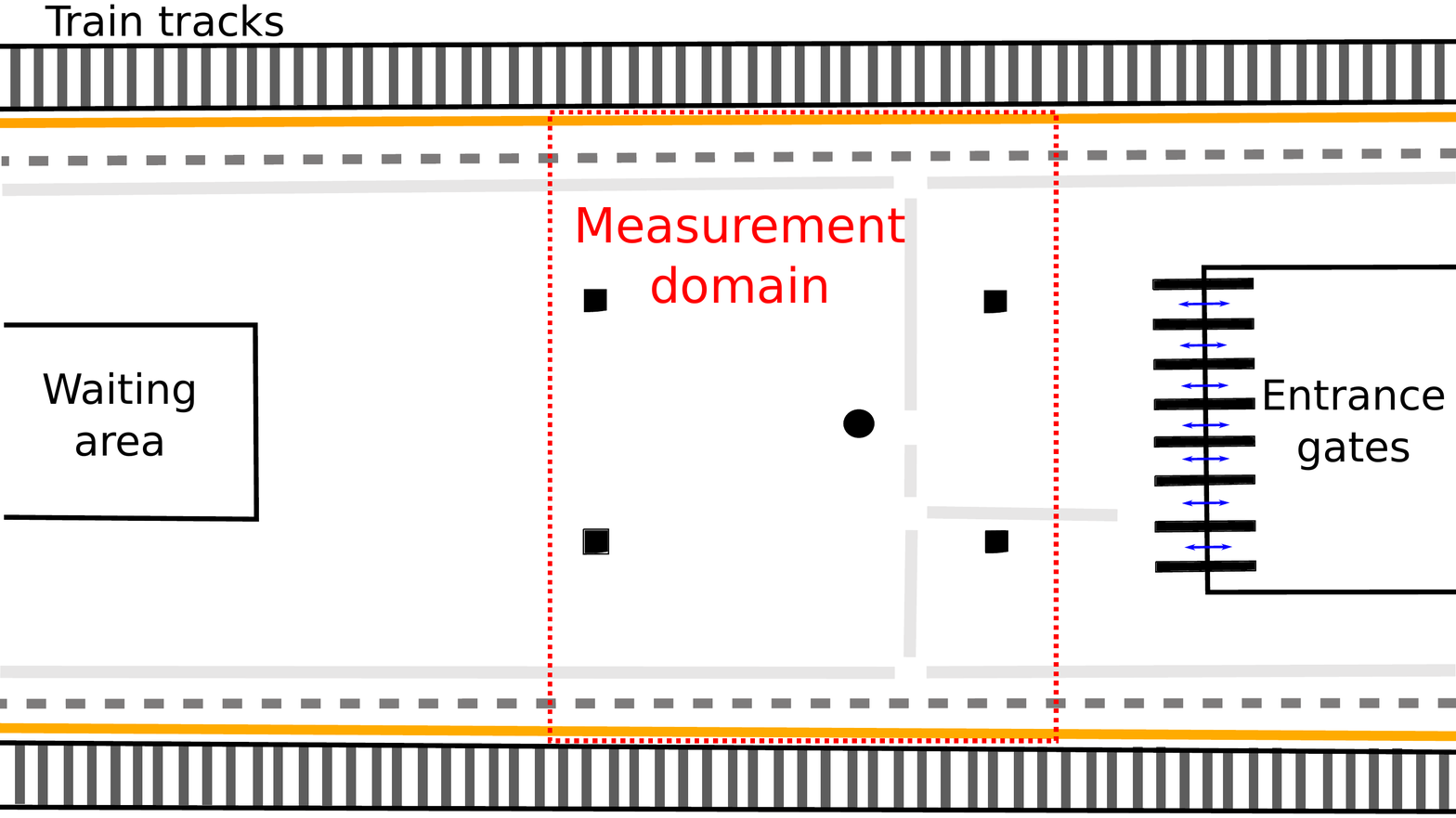}
  \caption{Experimental setup for the commercial tracking system at
    train station Breukelen, The Netherlands, visualized with a
    schematic floorplan. The train tracks are shown in the top and
    bottom of the image. The dimensions of the measurement domain,
    with size $4\;\textrm{m}\times12.4\;\textrm{m}=50\;\textrm{m}^2$, is highlighted with a red dashed
    rectangle.} 
  \label{fig: setup2}
\end{figure}

\paragraph{Commercial setup} The commercial system anonymously tracks
pedestrian movements using 3D stereoscopic images. The system consists
of 3 commercial pedestrian tracking sensors
(Xovis\textsuperscript{TM}), used in e.g. train
stations~\cite{Pouw2020, Heuvel2019, Thurau2019, Thurau2020}, to
monitor complex crowd flows. Every sensor records images at $f = 10$
frames per second and processes the stereo images in real-time only
storing pedestrian locations as x, y coordinate pairs. This system is
installed at real-life operational train station Breukelen, the
Netherlands. The sensors are mounted to the ceiling of a platform
covering an area of
$4\, \textrm{m} \times 12.4\, \textrm{m} \approx 50
\,\textrm{m}^2$. We report an overview of the platform in
Fig.~\ref{fig: setup2}.

Note that (most) commercial systems only return trajectories, whereas
in custom setups one can retain also the raw data which can be used to
further improve the algorithms and/or extracting additional
features such as body orientations.


\section{Pedestrian tracking benchmark}\label{sec: benchmark}
We describe here the five tests comprised in our benchmark, which is
also resumed and schematized in Table~\ref{tab: all-tests}
encompassing an illustration and key features for the individual
tests. We additionally present advises on test duration, $\Delta T$
and number of participants, $N$, using a gray cell background.
\paragraph{Test 1: Line-based crowd flux estimation.}
We probe the accuracy in estimating crowd fluxes, as count of
pedestrians crossing a line in a predefined time window, $t_0<t<t_1$. We compare
counts automatically estimated, $N(t_0,t_1)$, with the ground-truth,
$N_{real}(t_0,t_1)$, manually evaluated. As a final score we retain the
following indicator
\begin{equation}\label{eq: exp1}
  A^{(1)} = \left( 1-\frac{|N(t_0,t_1)-N_{real}(t_0,t_1)|}{N_{real}(t_0,t_1)} \right) \cdot 100,
\end{equation}
which equals $100$ in case of a correct count estimate and is lower
otherwise. 
Commercial systems generally include internal algorithms to compute
line crossings, whereas in the TU/e system we employ the algorithm
described in Appendix~\ref{app: counting_algo}. To reach a challenging
flux of $J_A \approx 100\; \textrm{ped/min}$, we ask participants to
follow a circular path, which is crossed twice by the straight line
across which the flux estimation occurs. In particular, we employ $N =12$
participants walking a loop with a diameter $D \approx 3\;$m for
$\Delta T \approx 5$ minutes.

\paragraph{Test 2: Local density estimation.} We target the accuracy in
estimating local pedestrian densities. We consider the number of
people $N_1,N_2,N_3,\ldots$ moving freely within virtual regions
$S_1,S_2,S_3,\ldots$ determined by the tracking systems, and compare
it with the ground-truth. For simplicity, we keep the number of
pedestrians in each virtual region constant, $N_{i, real}$, and
consider the following time-averaged relative error as the score
\begin{equation}\label{eq: exp2}
  A_i^{(2)} = \left( 1 - \frac{1}{T} \int^{T}_0 \frac{|N_i(t)-N_{i,real}|}{N_{i,real}}\,dt\right) \cdot 100.
\end{equation}
With few participants we only target low average crowd densities in
this test. However, the localization task is very challenging due to
short distances between first neighbors that yield instantaneously
high densities.
Additionally, (stationary) objects can be added to the test area to
validate the ability to differentiate between objects and people.

\paragraph{Test 3: Individual position detection.}
We target, in line with~\cite{Heuvel2019}, the capability of accurately determining individual
positions. To bypass the need to manually establish a ground-truth for
point-wise comparisons, we ask participants to walk following simple
geometric patterns, specifically, a grid of straight lines
(cf. markings in Fig.~\ref{fig: setup1}a).
We score how closely the measured trajectories agree, as an ensemble,
with the geometric pattern, i.e. they form thin and straight bands.

Operationally, for each collected trajectory, we isolate the portions
that follow single grid lines. For each grid line, we obtain a set of
trajectory pieces of which we consider averages. We either retain the
best fitting straight line (linear regression) or we find a piece-wise
average in bins of $D_{bin}=5\;$cm (local regression). Naming, without
loss of generality, these fitting curves, respectively,
$y_{lin} = y_{lin}(x)$ and $y_{loc} = y_{loc}(x)$, we quantify the
following (the segments naming is as in Tab.~\ref{tab: all-tests}
third entry):
\begin{itemize}
\item spread of trajectories along each grid line, that should be
  comparable to the individual pedestrians body sway amplitude (about
  $5\;$cm~\cite{Liu2009}). We consider for each $x$-location
  parametrizing a line the quantity $z_{lin}(x) = y_{lin}(x) - y(x)$
  where $y(x)$ is a generic measurement of coordinate $y$ at position
  $x$. Note that $z(x)$ is approximately zero-centered at each
  $x$. Our benchmark quantifies
\begin{equation}\label{eq: exp3}
  \sigma_{lin}^{(3)} =\underset{\textrm{at all}\;x}{\underset{\textrm{measurements}}{\textrm{std.dev.}}}(z_{lin}(x))
\end{equation}
likewise it holds for $\sigma_{loc}^{(3)}$;
\item distance between linear fits over parallel lines, that should be
  constant. We score them with slopes, such as the following, for the
  case of segments AK, BL (for the other segments the formula works
  similarly)
\begin{equation}\label{eq: exp3d}
D^{(3)} = 100 - \left( \frac{|DE| - |HI|}{|DH|}\right) \cdot 100;
\end{equation}
\item angles between linear fits over perpendicular lines, that should
  be $90\degree$. For the angle
  $\angle DHI$ we score this as follows (generalization for the other
  angles is not reported)
\begin{equation}\label{eq: exp3l}
  L^{(3)} =  100 - \left(\frac{|\angle DHI - 90\degree|}{90\degree}\right) \cdot 100.
\end{equation}
\end{itemize}

\paragraph{Test 4: Trajectory accuracy in controlled environment}
We target the capability of tracking pedestrians for extended time
periods and along complex trajectories. We divide the measurement
domain in regions $S_1,S_2,S_3,\ldots$, and ask each participant,
$id$, from a set of $N_{real}$ pedestrians, to stand in a region
$S_{o}^{id}$ and walk to an assigned destination $S_{d}^{id}$,
following an irregular path of choice taking roughly
$\Delta T \approx 30\;$s. Origin and destination regions are assigned
exclusively to a single pedestrian.  With few participants,
e.g. $N_{real}\approx 12$, the average density is low during this
test. However, instantaneously we have extremely high densities due to
short distances between first neighbors, this makes tracking very
challenging. A recorded trajectory is considered correct when its
origin and destination respectively lie within the boundaries of the
assigned origin-destination pair $(S_o^{id},S_d^{id})$. The final
score is the percentage of correct trajectories. To increase the
difficulty, (stationary) pedestrians standing outside all the regions
$S_1,S_2,S_3,\ldots$, can be added to the measurement domain.
\begin{equation}\label{eq: exp4}
A^{(4)}=\left(\frac{N_{corr}}{N_{real}}\right) \cdot 100
\end{equation}

\paragraph{Test 5: Trajectory accuracy in real-life
  environment}
Finally, we test in a real-life environment the capability of tracking
pedestrians without interruptions.
We define an inner region, $S_{in}$, in which
no trajectory can physically start or terminate.
Each trajectory recorded in a time window $\Delta T \geq
1\;$day, that enters the domain $S_{in}$, is classified according to
its quality:
\begin{enumerate}
\item \textbf{correct}: neither the trajectories initial nor final
  point lay inside region $S_{in}$;
\item \textbf{faulty termination}: the trajectory terminates
  inside region $S_{in}$;
\item \textbf{faulty origin}: the trajectory originates inside region $S_{in}$.
\end{enumerate}
\noindent
We report the percentage of trajectories correctly tracked $A{(5)}$,
approximating the total number of trajectories as the correct
trajectories plus broken trajectories. In formulas this reads
\begin{equation}\label{eq: exp5}
  A^{(5)} = \left( \frac{\textrm{correct}}{\textrm{correct} + \textrm{broken}} \right) \cdot 100,
\end{equation}
where broken trajectories are interrupted paths consisting of two or more trajectory
pieces. Because one trajectory piece must enter domain $S_{in}$ and another must leave
this domain we can approximate broken trajectories as
\begin{equation}
\textrm{broken} = \frac{\textrm{faulty termination} + \textrm{faulty origin}}{2}.
\end{equation}
\input{exp_table/experiments-table-portrait.tex}
\begin{figure}[p]
  \centering
  \includegraphics[width=0.53\linewidth]{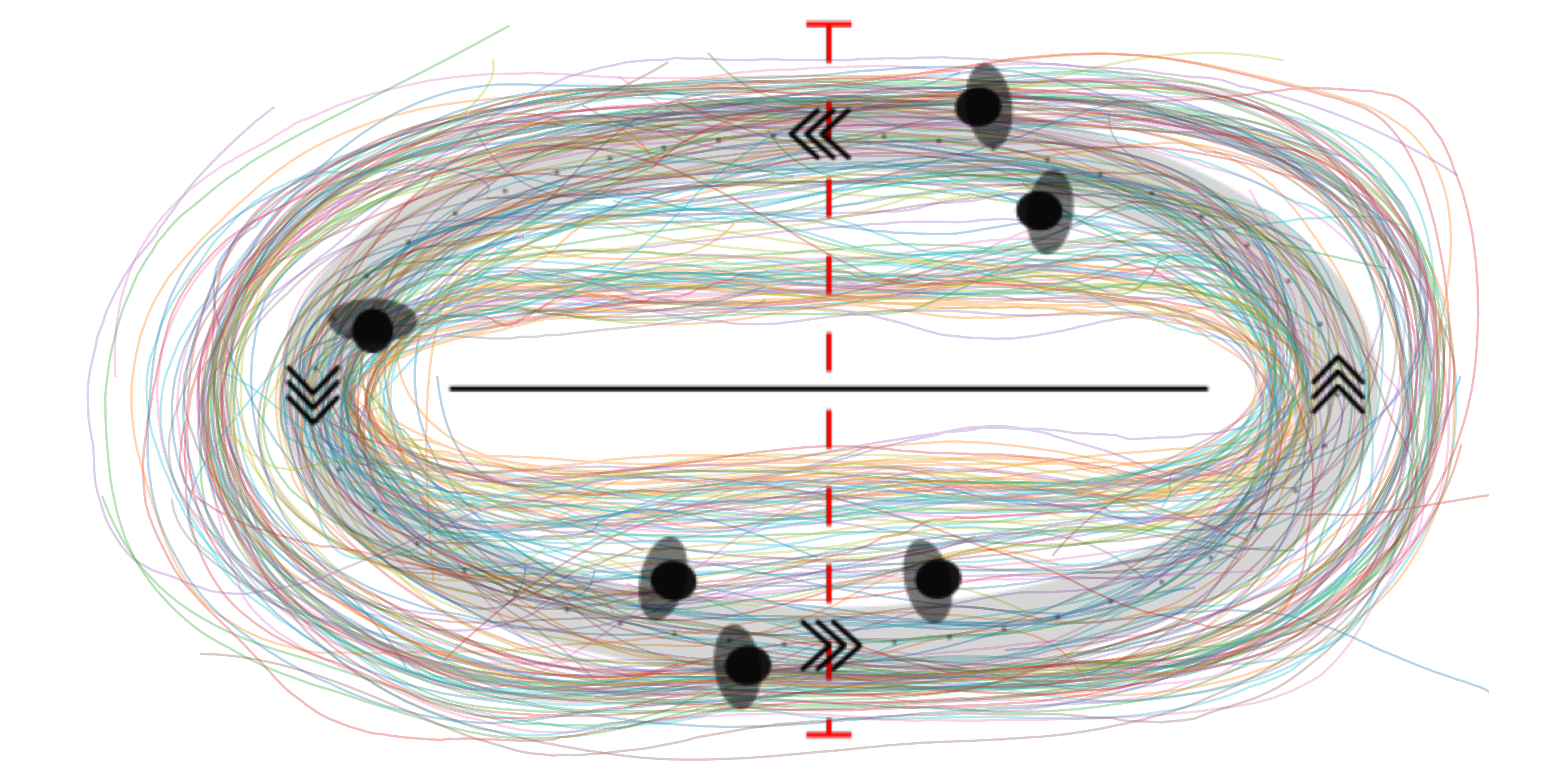}
  \includegraphics[width=0.45\linewidth]{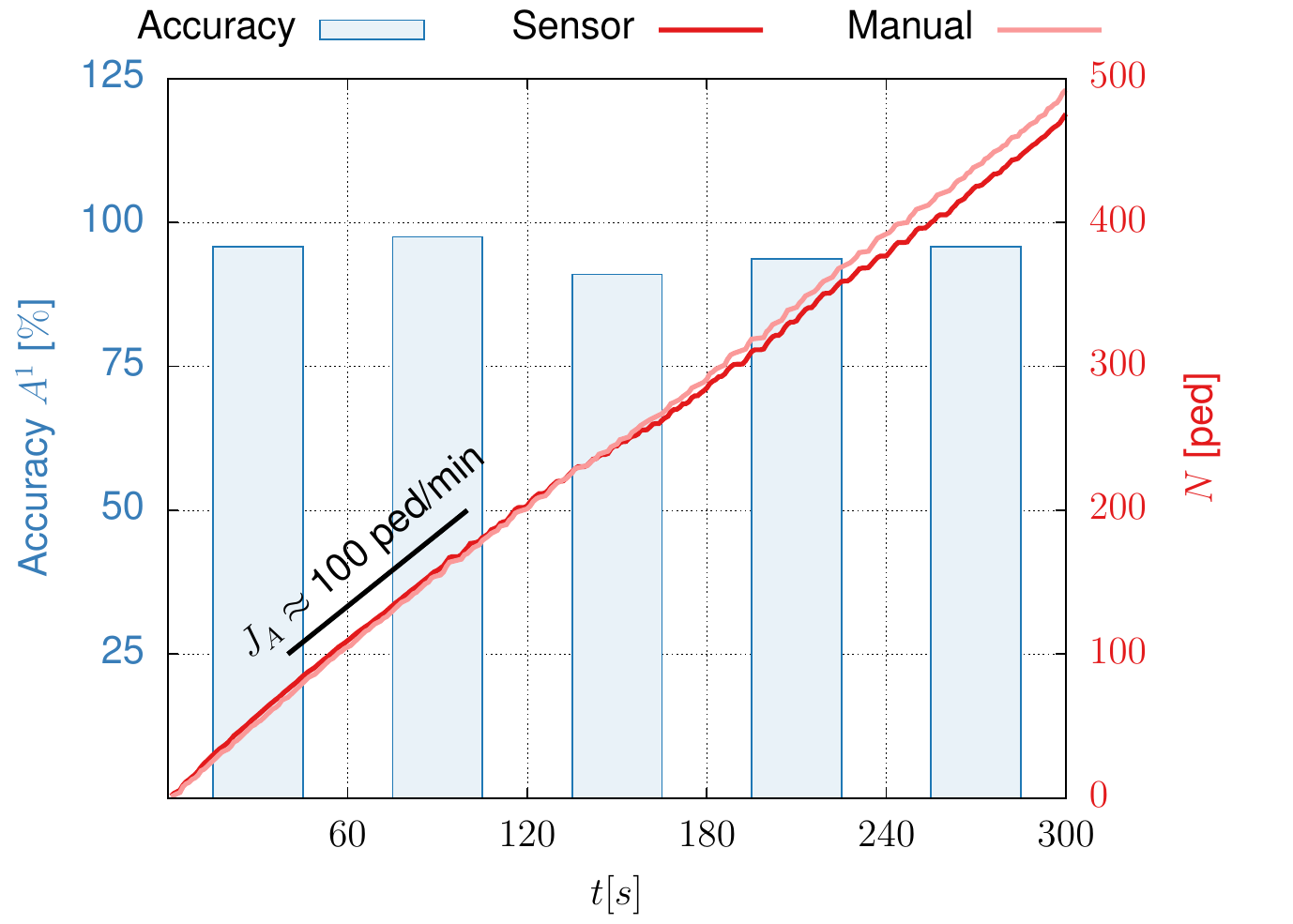}
  \caption{(a) Trajectories captured during test 1 by the TU/e
    setup, superimposed to the schematic illustration from
    Tab.~\ref{tab: all-tests}. (b) Results of test 1 for the TU/e
    setup. Blue bars report the accuracy, $A^{(1)}$, on a
    minute-by-minute basis, red lines indicate measured (dark) and
    ground-truth (light) pedestrian count, $N(t)$, and a black line
    indicates the slope i.e. the average crowd flux,
    $J_A \approx 100\; \textrm{ped/min}$.}
  \label{fig: LC}
\end{figure} 
{\renewcommand{\arraystretch}{1.7}
  \begin{table}[p]
    \centering
  \begin{tabular}{l || l | l | l || l | l | l}
     & \multicolumn{3}{c||}{\textbf{TU/e}} & \multicolumn{3}{c}{\textbf{Commercial}} \\
    \multirow{2}{*}{Minute} & \multicolumn{3}{l||}{$J_A \approx 100$ ped/min} & \multicolumn{3}{l}{$J_A \approx 30$ ped/min} \\
     & $N_{real}$ & $N-N_{real}$ & $A^{(1)}$ & $N_{real}$ & $N-N_{real}$ &  $A^{(1)}$ \\ \hline \hline
    1 & 105 & 4 & 96$\%$ & 17 & 0 & 100$\%$ \\ \hline
    2 & 96 & 2 & 98$\%$& 38 & 0 & 100$\%$ \\ \hline
    3 & 90 & 8 & 91$\%$ & 37 & 0 & 100$\%$ \\ \hline
    4 & 98 & 6 & 94$\%$ & 29 & 0 & 100$\%$ \\ \hline
    5 & 103 & 4 & 96$\%$ & NA & NA & NA \\ \hline \hline
    Total & 492 & 24 & 95$\%$ & 121 & 0 & 100$\%$
  \end{tabular}
  \caption{Synthetic results of test 1 for both pedestrian tracking
    setups. We report, on a minute-by-minute basis, the ground-truth
    number of pedestrians, $N_{real}$, the error in the count,
    $N - N_{real}$, and the estimation accuracy, $A^{(1)}$. While
    testing the TU/e setup, we considered a crowd flux of
    $J_A = 100\; \textrm{ped/min}$, whereas in the
    commercial case, a far less challenging crowd flux of
    $J_A = 30\; \textrm{ped/min}$ was maintained. Therefore, the test
    of the commercial system can be considered relatively less difficult.  
  }
  \label{tab: exp1}
\end{table}}
\section{Benchmark results on the considered systems}\label{sec: results}
\begin{figure}[p]
  \centering
  \includegraphics[trim = 0 100 0 130, clip,
    width=0.48\linewidth]{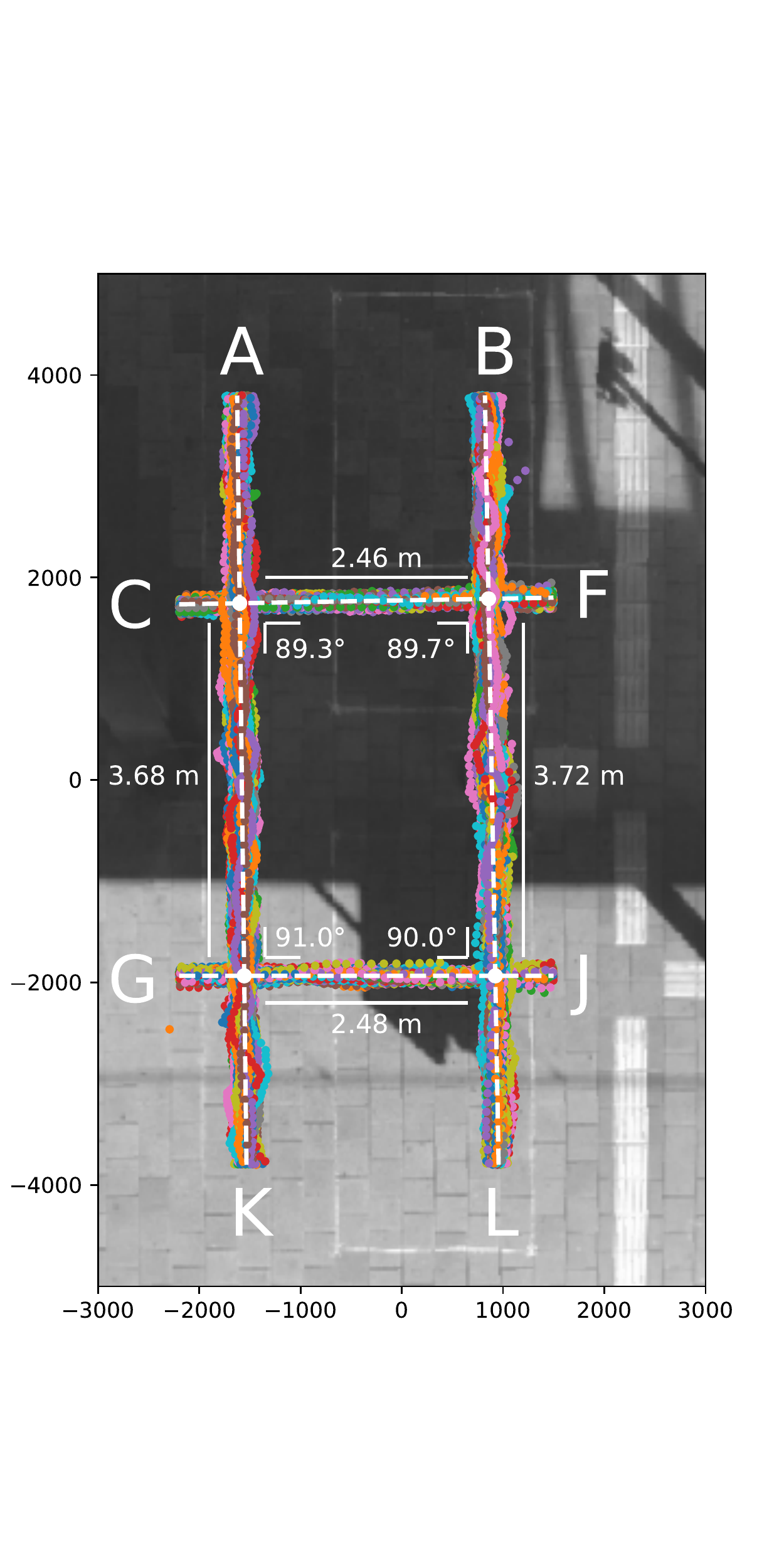}
  \includegraphics[width=0.25\linewidth]{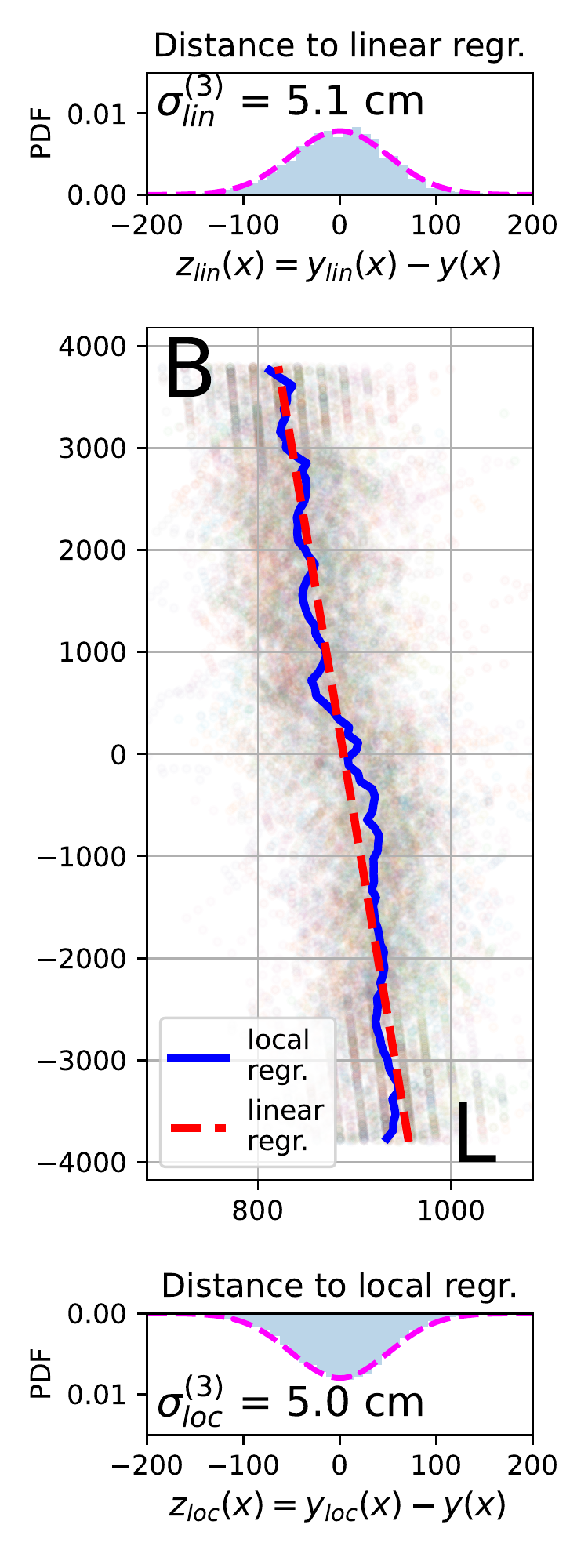}
  \caption{(a) Recorded trajectories by the commercial setup during
    test 3. The figure reports a linear fit for every set of trajectory
    pieces. The regressions reconstruct the geometric structure followed by the
    participants. Additionally, we report the distances
    between parallel fits and the angles between perpendicular
    fits. (b) Isolated trajectory pieces for grid line BL. We fitted
    a linear (blue) and local (red) regression through the trajectory
    pieces. Additionally, we report the histograms of $z_{lin}(x)$
    (top) and $z_{loc}(x)$ bottom including a Gaussian fit (pink).
  }
  \label{fig: exp3}
\end{figure}

\begin{figure}[p]
  \centering
   \includegraphics[width=0.25\linewidth]{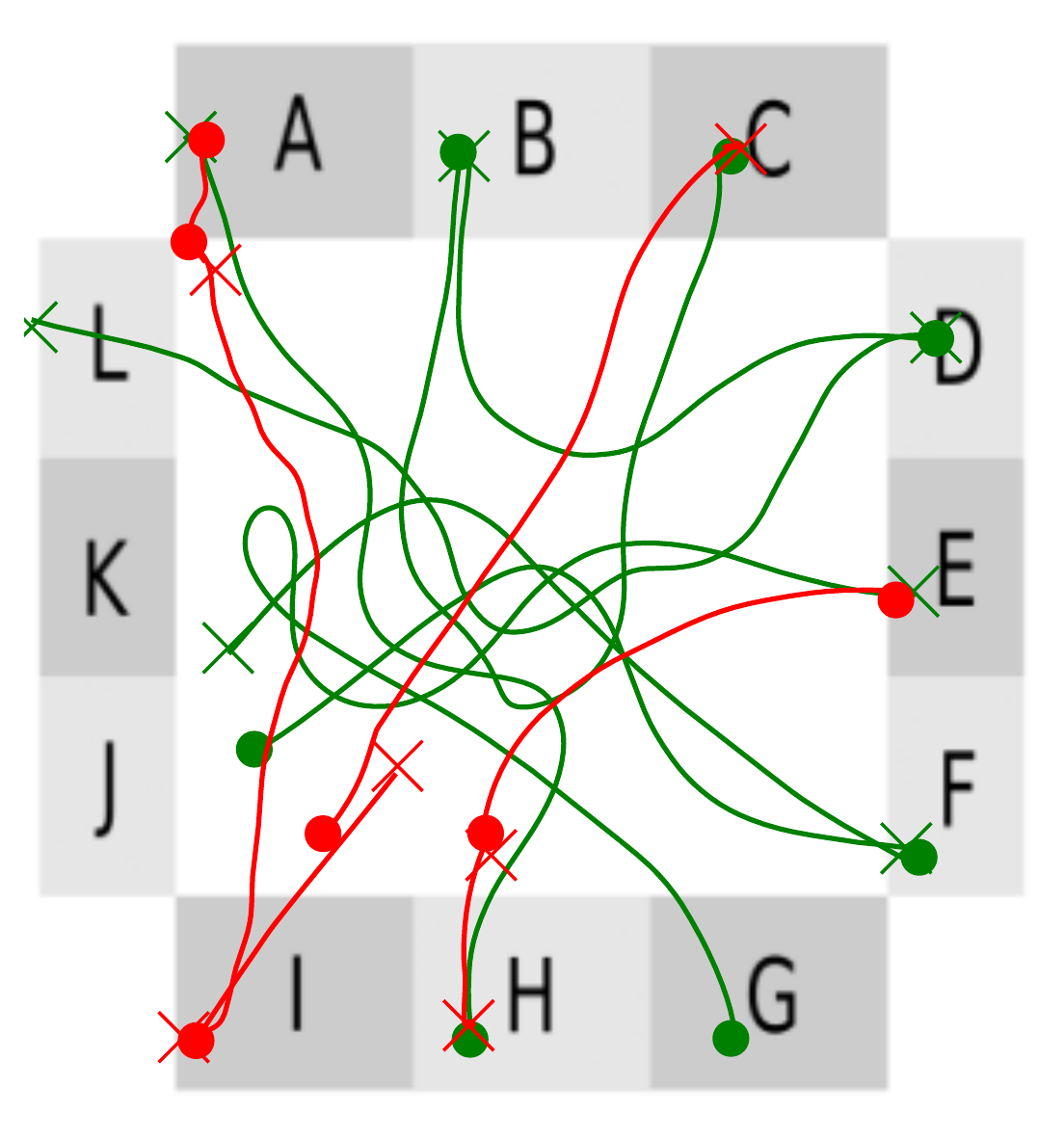}
  \includegraphics[width=0.25\linewidth]{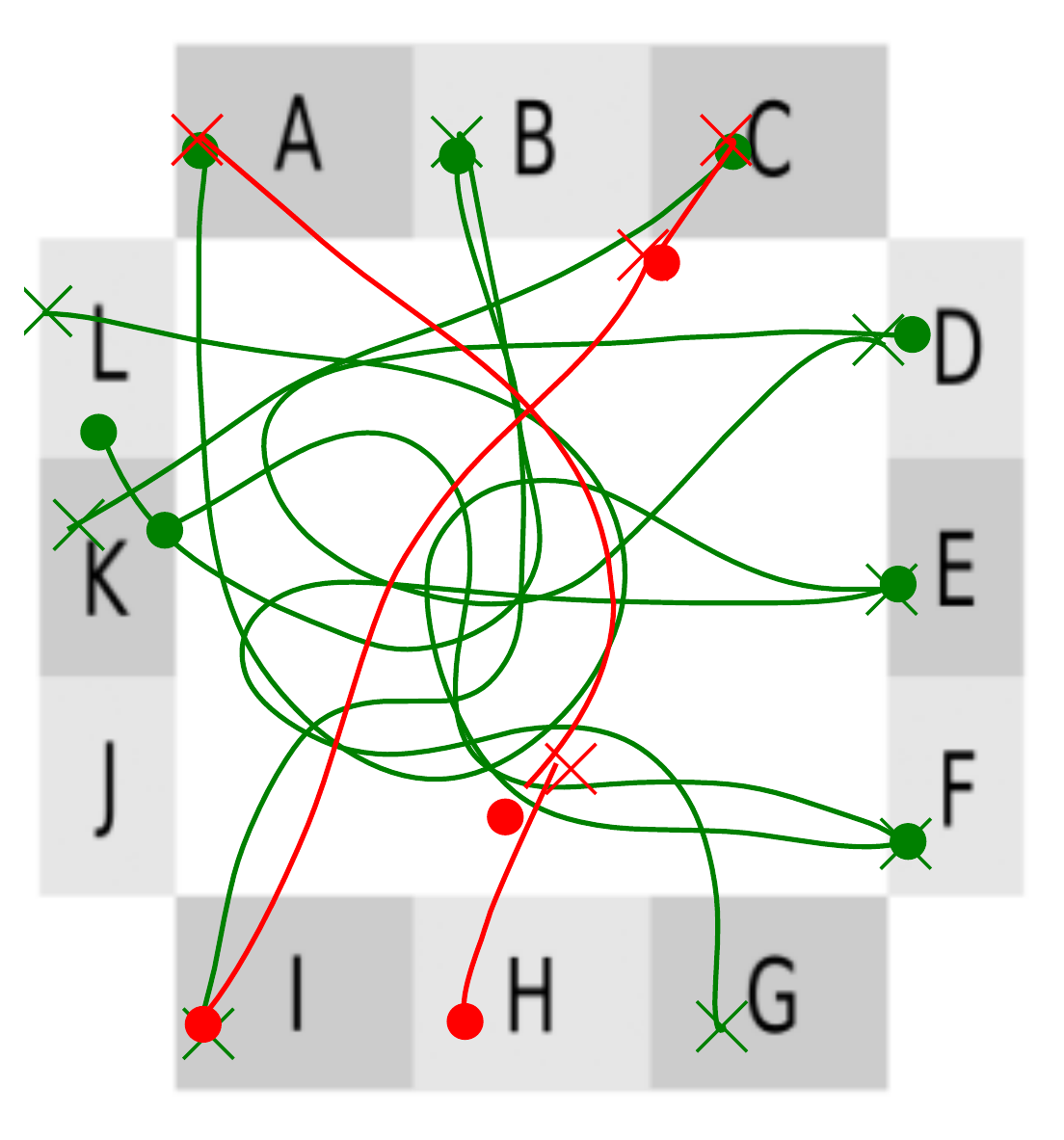}
  \includegraphics[width=0.25\linewidth]{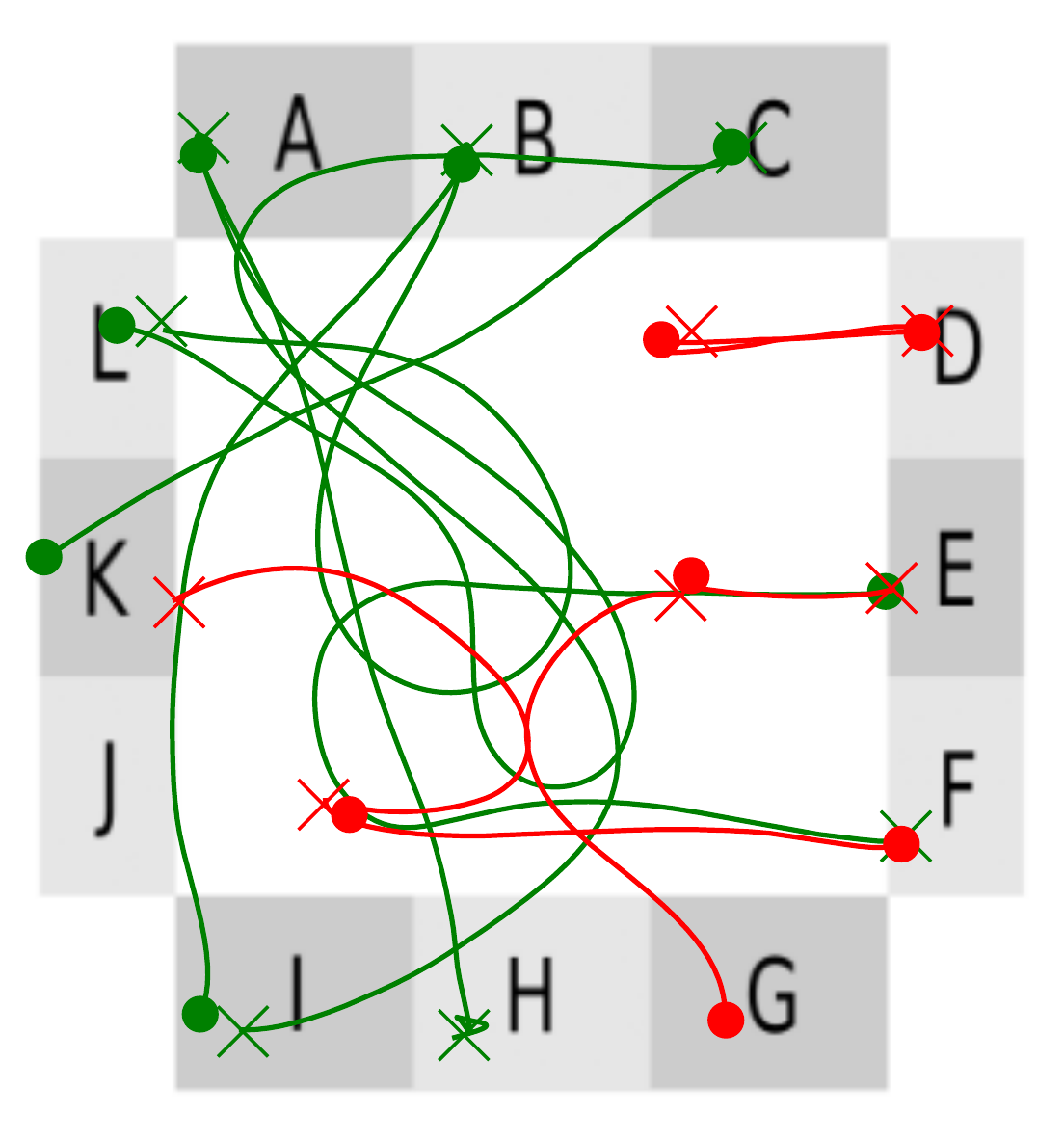}
  \caption{Trajectories captured by the TU/e setup during test
    4. Green lines indicate correct trajectories whereas red lines
    indicate faulty trajectories. Trajectory origins and destinations
    are indicated with spheres and crosses respectively.
  }
  \label{fig: exp4}
\end{figure}

\begin{figure}[t]
  \centering
  \includegraphics[width=.8\linewidth]{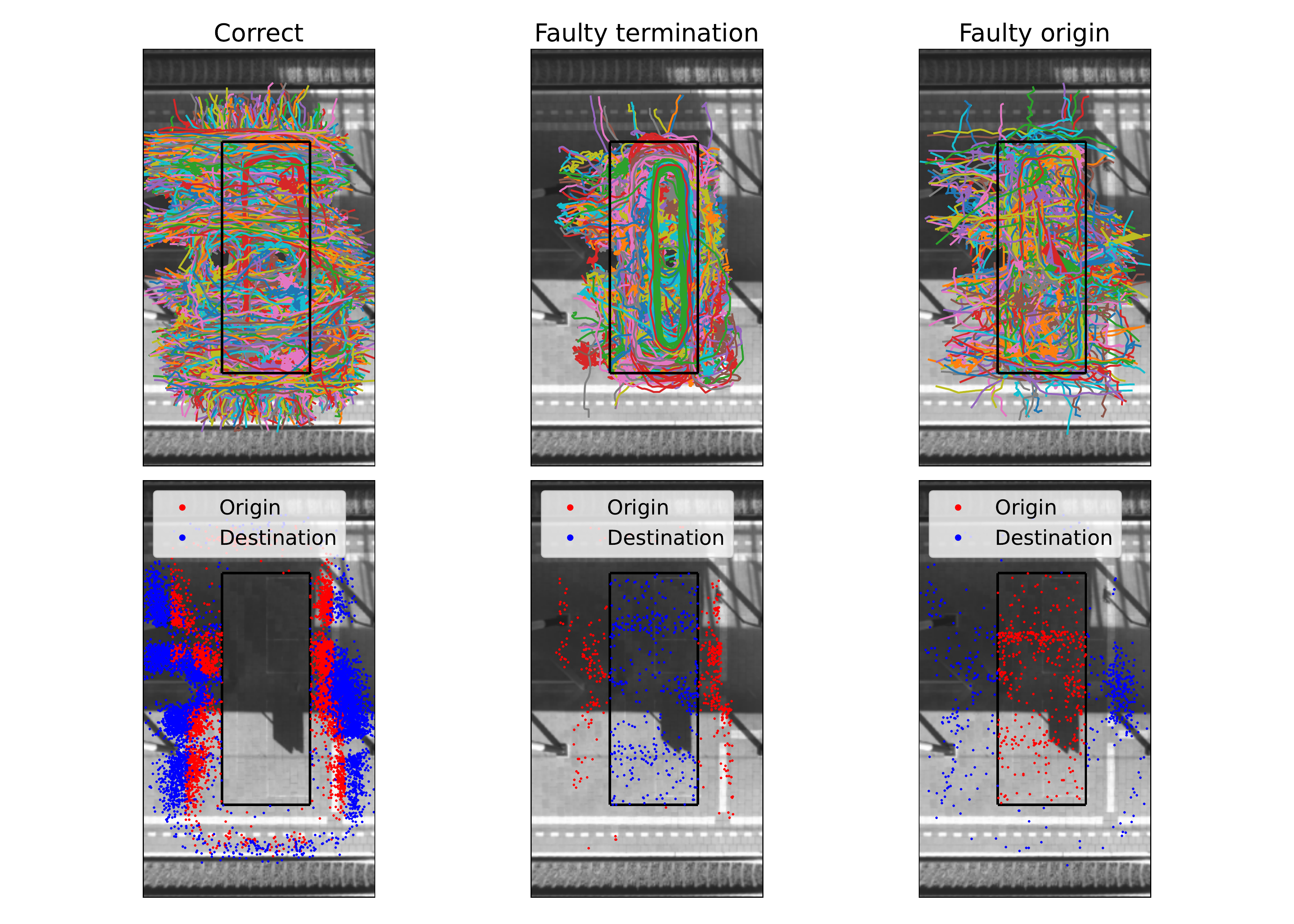}
  \caption{
    The recorded trajectories by the commercial setup during test
    5. The upper row with figures reports the trajectories, partitioned
    per subset, and the bottom row reports the
    corresponding origin-destinations pairs. The black rectangle indicates the
    inner domain, $S_{in}$.
  }
  \label{fig: TCRL}
\end{figure}

In this section we report and elaborate on the benchmark results for
the pedestrian tracking systems introduced in Section~\ref{sec:
  setups}. We iteratively improved each test by trying different
methods and variants. This optimization process, in combination with
the ever-changing nature of the real-life testing environments,
caused, for some tests, minor differences between the two setups. For
each test, we report synthetic results plus illustrations taken
from either setups \\[.2cm]
\textbf{Test 1: Line-based crowd flux estimation.}  In Fig.~\ref{fig:
  LC} we report a sample of captured trajectories and the results, in
graph form, for the TU/e measurement setup. Specifically, in
Fig.~\ref{fig: LC}b we report the minute-by-minute estimation accuracy
(blue bars), the cumulative pedestrian count (red line), and the
average crowd flux (black slope). Synthetic results for both setups
are in Tab.~\ref{tab: exp1}, which includes, in time windows
$t_0<t<t_1$ of $1$ minute, the ground-truth pedestrians count,
$N_{real}(t_0,t_1)$, the crowd flux estimation error,
$N(t_0,t_1)-N_{real}(t_0,t_1)$, and the estimation accuracy, $A^{(1)}$
(Eq.~\ref{eq: exp1}). The commercial setup is tested with an average
crowd flux of $J_A\approx 30\;\textrm{ped/min}$, whereas the TU/e
measurement setup is tested with a more challenging
$J_A\approx 100\;\textrm{ped/min}$. Results of the TU/e and
commercial setups are $A^{(1)}=95\%$ and $A^{(1)}=100\%$ respectively,
however the relative difficulty of the commercial setup can be
quantified to $\sim 1/3$ due to the
lower crowd flux. \\[.2cm]
\noindent \textbf{Test 2: Local density estimation.}  We report, for
the commercial setup, the test layout in Fig.~\ref{fig: LC}a and the
test results in Fig.~\ref{fig: LC}b. For the commercial setup we
define two regions $S_1,S_2$ (cf. Fig.~\ref{fig: LC}a), both with an
area of $6.2\;\textrm{m}^2$. In the case of the TU/e setup we employed
only one larger region $S=150\;\textrm{m}^2$. To improve test
reliability, we perform multiple runs for each setup, thereby
realizing four density estimations each $A-D$, see Tab.~\ref{tab:
  exp2}. In the table we report also the ground-truth region
occupation, $N_{real}$, the local density, $\rho=\frac{N_{real}}{S}$,
and the estimation accuracy $A^{(2)}$ (Eq.~\ref{eq: exp2}). The
commercial setup shows a systematic error in the density estimation of
region $S_1$, most likely related to false positive detection of $2$
stationary objects. Therefore, in Tab.~\ref{tab: exp2}, we include an
additional column containing a corrected estimate $A_c^{(2)}$, for
which the systematic error is removed by reducing the pedestrian
count, $N(t)$, by $2$.
\begin{figure}[t]
  \centering
  \includegraphics[width=0.25\linewidth]{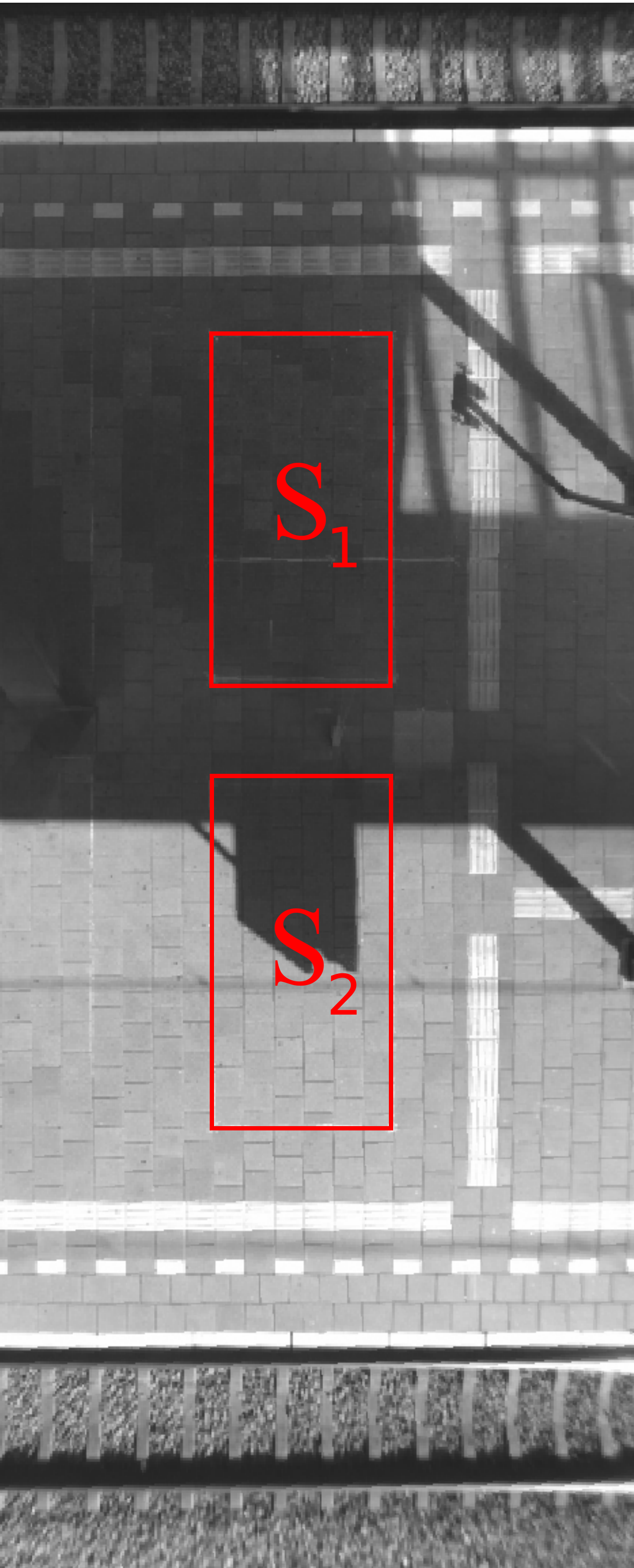}
  \includegraphics[width=0.64\linewidth]{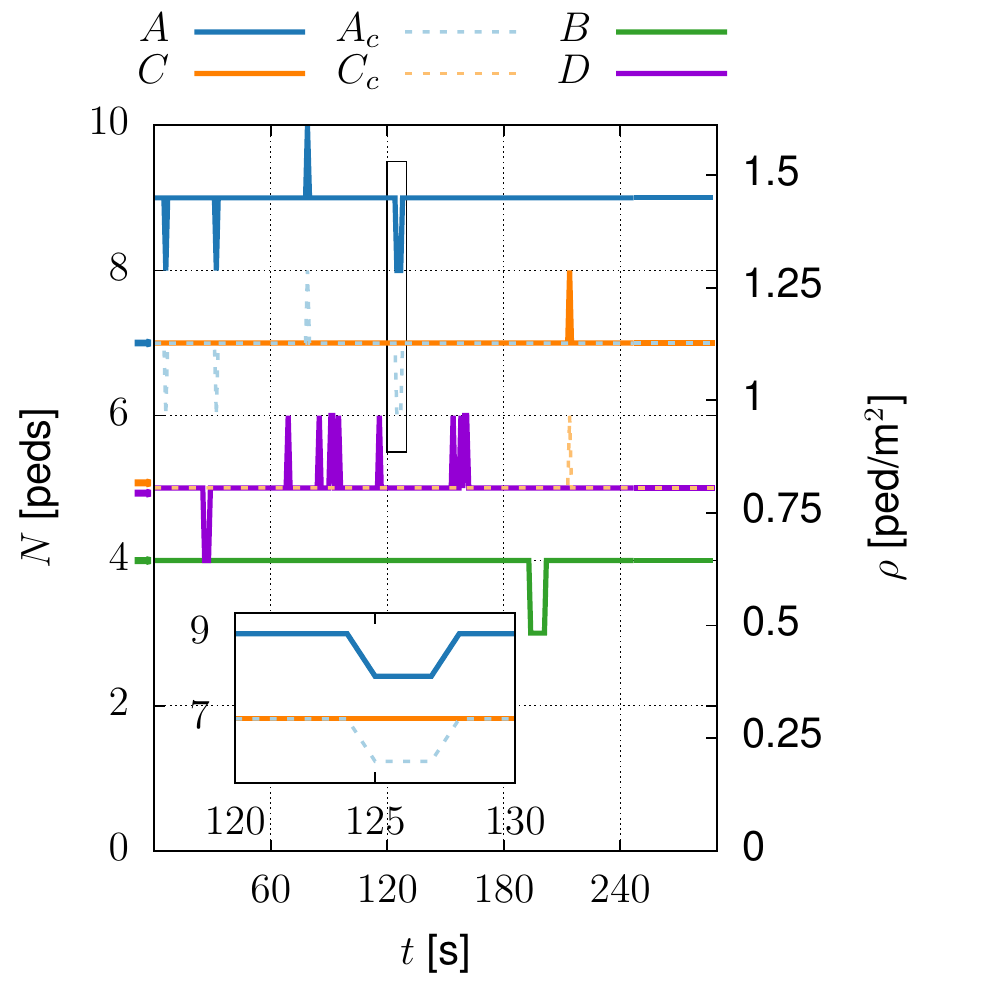}
  \caption{(a) Measurement domain for the commercial
   setup. Red domains indicate regions $S_1$ and $S_2$ both with an
   area $6.2\;\textrm{m}^2$.
   (b) Results of test 2 for the commercial setup. The
   figure reports for all repetitions, $(A-D)$, the measured
   pedestrian count, $N(t)$. A color matching mark on the left y-axis
   indicates the ground-truth pedestrian count $N_{real}$ for each repetition.
   Additionally, we report for $A$ and $C$ a systematic
   error corrected pedestrian count denoted $A_c$ and $C_c$
   respectively. The left y-axis indicates pedestrian
   count whereas the right y-axis shows the corresponding crowd density
   $\rho$ inside the $6.2\;\textrm{m}^2$ region.
    The inset shows an enlarged view of the graph part inside the black box.
  }
  \label{fig: LC}
\end{figure}
{\renewcommand{\arraystretch}{1.7}
  \begin{table}[t]
    \centering
  \begin{tabular}{l || l | l | l || l | l | l | l }
    & \multicolumn{3}{l||}{\textbf{TU/e}} & \multicolumn{4}{l}{\textbf{Commercial}} \\
    \multirow{2}{*}{Test} & \multicolumn{3}{l||}{$S = 150 \; \textrm{m}^2 \quad \Delta T \approx 300\; \textrm{s}$ } & \multicolumn{4}{l}{$S = 6.2\; \textrm{m}^2  \quad \Delta T = 300\; \textrm{s}$} \\
     & $N_{real}$ & $\rho\;[\textrm{ped/m}^2]$ & $A^{(2)}$ & $N_{real}$ & $\rho\;[\textrm{ped/m}^2]$ & $A^{(2)}$ & $A_c^{(2)}$ \\ \hline \hline
    A & 10 & 0.07 & $99\%$ &  7 & 1.13 & $71\%$ & $100\%$ \\ \hline
    B & 8 & 0.05 & $99\%$ & 4 & 0.65 & $99\%$ & $99\%$ \\ \hline
    C & 11 & 0.07 & $98\%$ & 5 & 0.81 & $60\%$ & $100\%$ \\ \hline
    D & 12 & 0.08 & $94\%$ & 5 & 0.81 & $99\%$ & $99\%$ \\ \hline \hline
    Total & 41 & 0.07& $97\%$ & 21 & 0.85 & $83\%$ & $100\%$
  \end{tabular}
  \caption{Synthetic results of test 2 for both tracking setups. We
    report for the repetitions $A-D$ the ground-truth number of pedestrians,
    $N_{real}$, the local density, $\rho$ and the estimation error,
    $\epsilon^{(2)}$. The TU/e setup estimates density over an area
    of $S=150 \textrm{m}^2$ whereas the commercial setup considers
    much smaller regions of $S=6.2 \textrm{m}^2$. Due to the lower
    crowd density the test of the TU/e setup can be considered
    relatively a factor 10 less difficult. For the commercial setup we report
    additionally the estimation error after correction for
    systematic errors $A_c^{(2)}$.
  }
  \label{tab: exp2}
\end{table}  }
Among all four tests the commercial setup sustained an average local
density of $\rho = 0.85\;\textrm{ped/m}^2$, whereas the density across
the TU/e measurement domain is a factor $10$ lower. Because
pedestrian localization is more challenging in dense crowds, we
reflect the difference in pedestrian density in the relative
difficulty. Throughout run $D$, the infrared sensor of the TU/e
setup is overexposed by excessive sunlight. Reduced image quality
causing false negatives results, for this run, in a lower density estimation
accuracy.\\[0.2cm]
\noindent \textbf{Test 3: Individual position detection.}  We report in
Figure~\ref{fig: exp3}a, for the commercial setup, the recorded
trajectories during test 3. We fit a linear regression for every
isolated set of trajectory pieces. The regressions accurately
reconstruct the geometric structure that is followed by the
participants i.e. closely resemble a grid of straight lines. We added
the angles and distances between the grid lines to emphasize the
correspondence. Figure~\ref{fig: exp3}b provides the isolated set of
trajectory pieces belonging to grid line BE. We fit a local (blue) and
linear (red) regression through the trajectory pieces. Additionally,
we report the spread along the grid line with histograms for
$z_{lin}(x)$ (top) and $z_{loc}$ (bottom) annotated with test scores
$\sigma_{lin}^{(3)}$ and $\sigma_{loc}^{(3)}$. We refer to
Table~\ref{tab: exp3} for the standard deviations,
$\sigma_{lin}^{(3)}$ and $\sigma_{loc}^{(3)}$, for both setups. All
standard deviations are the same order as typical body sway amplitude
i.e. $\sigma^{(3)} = 5\;$cm. The test is relatively more challenging
for the TU/e setup which needs more sensors for the large measurement
domain. In Table~\ref{tab: exp3ang} we report the correspondence to
the grid geometry in terms of the distance between trajectories on
parallel grid lines $D^{(3)}$, and the angle between trajectories on
perpendicular grid lines $L^{(4)}$. The grid geometry is reconstructed
with high accuracy as the angles and the mutual distances in the
recorded trajectories agree up to $99\%$ with the original grid
structure.
{\renewcommand{\arraystretch}{1.7}
  \begin{table}[p]
    \centering
  \begin{tabular}{l || l | l | l || l | l | l}
    \multirow{2}{*}{Test} & \multicolumn{3}{l||}{\textbf{TU/e}} & \multicolumn{3}{l}{\textbf{Commercial}} \\
     & $N_{sens}$  & $\sigma_{loc}^{(3)}$ [cm] & $\sigma_{lin}^{(3)}$ [cm] & $N_{sens}$ & $\sigma_{loc}$ [cm] & $\sigma_{lin}$ [cm] \\ \hline \hline
    AK & 4 & 6.0 & 8.7 & 3 & 5.0 & 5.1 \\ \hline
    BL & 4 & 4.9 & 5.5 & 3 & 5.1 & 5.2  \\ \hline
    CF & 3 & 3.0 & 5.6 & 1 & 3.8  & 3.8 \\ \hline
    GJ & 3 & 5.8 & 8.5 & 1 & 3.5 & 3.5 \\ \hline \hline
    Average & 3.5 & 4.9 & 7.1 & 2 & 4.4 & 4.4
  \end{tabular}
  \caption{Synthetic results of test 3 for both tracking
    setups. The table reports for each grid line, the number of
    overhead sensors, $N_{sens}$, and the standard
    deviation with respect to the local, $\sigma_{loc}$ and to the linear,
    $\sigma_{lin}$, regression.
  }
  \label{tab: exp3}
\end{table}  }
{\renewcommand{\arraystretch}{1.7}
  \begin{table}[p]
    \centering
  \begin{tabular}{l || l || l }
    Test & \textbf{TU/e} & \textbf{Commercial} \\ \hline \hline
    $D^{(3)}$ & $98.8\;\%$ & $98.8\;\%$ \\ \hline
    $L^{(3)}$ & $98.9\;\%$ & $99.4\;\%$ 
  \end{tabular}
  \caption{Synthetic results of test 3 for both setups.
    The table reports how accurate the recorded trajectories agree
    with the grid geometry. In particular, we report the accuracy in
    reconstructing parallel grid lines, $D^{(3)}$, and the accuracy in
    reconstructing perpendicular grid lines, $L^{(3)}$.
  }
  \label{tab: exp3ang}
\end{table}  } \\[0.2 cm]
\noindent \textbf{Test 4: Trajectory accuracy in controlled
  environment.}  In Figure~\ref{fig: exp4}, we report side-by-side the
trajectories recorded by the TU/e measurement setup during the three
runs of test 4. Correct trajectories have a green and faulty
trajectories a red color. Table~\ref{tab: exp4} reports the test
results for both setups indicating, for each run, the number of
trajectories, $N_{real}$, the number of correct trajectories,
$N_{corr}$, and the trajectory accuracy, $A^{(4)}$. The participants
experience high instantaneous densities with almost body-to-body
contact. Minimum mutual distances in the order $20\;$cm are recorded
several times over the lifespan of their trajectories. Some tests of
the commercial setup also contained additional stationary pedestrians to
increase the local density, this is represented in the table with an
extra column $N_{obj}$. The TU/e setup records, over an area of
$S = 150\;\textrm{m}^2$, $22$ correct from a total of $30$
trajectories whereas the commercial setup, on a much smaller area
$S = 50\;\textrm{m}^2$, scores $49$ out of $64$ trajectories. Both
setups report a trajectory accuracy in the order of
$A^{(4)}\approx75\%$ (Eq.~\ref{eq: exp4}). This shows that in
conditions with highly entangled trajectories tracking procedures can
be imperfect, as only $75\%$ of the trajectories are
captured properly. \\[0.2cm]
\noindent \textbf{Test 5: Trajectory accuracy in real-life environment.}
In Figure~\ref{fig: TCRL} we report all trajectories, recorded during
$\Delta T = 1$ day, partitioned in subsets: correct, faulty
termination, and faulty origin (cf. Sec.~\ref{sec: benchmark}) using an inner domain
$S_{in} \approx 38\;$m$^2$. The first
row of figures reports, per subset, the raw trajectories and the
second row reports all the origins (red) and destinations (blue) of
trajectories in the corresponding subsets. The percentage accurately
tracked trajectories, $A^{(5)}$, is determined to be $79\%$, which is in the same
order as the trajectory accuracy test in controlled
conditions. This shows that under normal operational conditions
$79\%$ of the trajectory recordings is interrupted and broken into
smaller pieces.
{\renewcommand{\arraystretch}{1.7}
  \begin{table}[p]
    \centering
  \begin{tabular}{l || l | l | l || l | l | l | l}
    \multirow{3}{*}{Test} & \multicolumn{3}{l||}{\textbf{TU/e}} & \multicolumn{4}{l}{\textbf{Commercial}} \\
     & \multicolumn{3}{l||}{\pbox{\linewidth}{$S = 150\; \textrm{m}^2$ \\ $d_{min}\approx 20\;$cm}} & \multicolumn{4}{l}{\pbox{\linewidth}{$S = 50\; \textrm{m}^2$ \\ $d_{min}\approx 20\;$cm}} \\
     & $N_{real}$& $N_{cor}$ & $A^{(4)}$ & $N_{real}$ & $N_{obj}$ & $N_{cor}$ & $A^{(4)}$ \\ \hline \hline
    A & 10 & 7 & 70$\%$ & 16 & 0 & 12 & 75$\%$ \\ \hline
    B & 10 & 8 & 80$\%$ & 16 & 0 & 12 & 75$\%$ \\ \hline
    C & 10 & 7 & 70$\%$ & 16 & 3 & 11 & 69$\%$ \\ \hline
    D & NA & NA & NA & 16 & 4 & 14 & 88$\%$ \\ \hline \hline
    Total & 30 & 22 & 73$\%$ & 64 & 7 & 49 & 77$\%$
  \end{tabular}
  \caption{Synthetic results of test 4 for both setups. We report for
    runs $A-D$ the number of participants, $N_{real}$, the number of correctly tracked trajectories, $N_{cor}$, and the
    test accuracy, $A^{(4)}$. Additionally we report for the
    commercial setup the number of
    (stationary) objects, $N_{obj}$.
  }
  \label{tab: exp4}
\end{table}  }
\section{Discussion}\label{sec: discussion}
In this contribution we presented a benchmarking suite for pedestrian
tracking systems. The suite is light-weight and easily reproducible as
it only contains a minimal set of 5 tests. The developed tests are
tailored to take minimal efforts, taking typically less than two hours
in total, while requiring only a dozen participants. Each test
accurately targets the validation of one of the following key
components of pedestrian tracking: line-based crowd flux estimation,
local density estimation, individual position detection, and
trajectory accuracy. The tests output quality factors expressed as
single numbers. The combination of tests focuses on error-prone
features like person-object recognition, and multi-sensor stitching.
From a civil engineering standpoint, the tests reflect observables
connecting with immediate awareness of a facility (1. instantaneous
usage, 2. crowding distribution), as well as with longer-term
efficiency and design (3. localization, 4-5. tracking, i.e. usage
modes). Facility usage and crowd distributions can indicate
potentially hazardous capacity issues and overcrowding in an early
stage, while localization and tracking enable efficiency improvements
such as separation of usage mode.

\bigskip

Together with the benchmarking suite we presented the benchmark
results of two real-life pedestrian tracking systems, one commercial
and one developed in academia. These test results, synthesized in
Tab.~\ref{tab: results}, are meant as a reference of the
state-of-the-art for new tracking installations and as a standard for
novel tracking technologies. The higher accuracy of the commercial
setup can be easily explained by its smaller measurement setup using
fewer sensors. The high error in the density estimation test for the
commercial setup is most likely caused by a systematic error due to
faulty person-object differentiation. This emphasizes the great
importance of background removal and proper sensor calibration. The
benchmark results show us that optic-based tracking systems can
estimate crowd fluxes and local densities, and localize pedestrians
with high accuracy. The biggest open challenge is Lagrangian
time-tracking in case of complex and highly intertwined trajectories
by pedestrians walking in close proximity. In this case the systems
scored an accuracy of about $75\%$.
{\renewcommand{\arraystretch}{2}
  \begin{table}[h]
    \centering
  \begin{tabular}{c | c | c | c | c}
    \textbf{Test} & \textbf{Name} & \textbf{Metric} & \textbf{TU/e setup} & \textbf{Commercial setup} \\ \hline \hline
    \textbf{1.} & \pbox{5cm}{Line-based crowd flux \\ estimation} & $A^{(1)}$ & \pbox{\linewidth}{$95\%$ \\ \scriptsize{at $J_A=100$ ped/m}} & \pbox{\linewidth}{$100\%$ \\ \scriptsize{at $J_A=30$ ped/m}} \\ \hline
    \textbf{2.} & \pbox{5cm}{Local density \\ estimation} &  $A^{(2)}$ & \pbox{\linewidth}{$97\%$ \\ \scriptsize{at $\rho = 0.07$ ped/m$^2$}} & \pbox{\linewidth}{$82\%$ \\ \scriptsize{at $\rho = 0.85$ ped/m$^2$}} \\ \hline
    \multirow{4}{*}{\textbf{3.}} & \multirow{4}{*}{\pbox{5cm}{Individual position \\ detection}} &  $\sigma_{lin}^{(3)}$ & \pbox{\linewidth}{$7.1\;$cm \\ \scriptsize{with $N_{sens} = 12$}} &  \pbox{\linewidth}{$4.4\;$cm\\ \scriptsize{with $N_{sens} = 3$}} \\ \cline{3-5}
    & & $\sigma_{loc}^{(3)}$ &  \pbox{\linewidth}{$4.9\;$cm \\ \scriptsize{with $N_{sens} = 12$}} &  \pbox{\linewidth}{$4.4\;$cm \\ \scriptsize{with $N_{sens} = 3$}} \\ \cline{3-5}
    & & $D^{(3)}$ & \pbox{\linewidth}{$98.8\%$ \\ \scriptsize{with $N_{sens} = 12$}}  & \pbox{\linewidth}{$98.8\%$ \\ \scriptsize{with $N_{sens} = 3$}} \\ \cline{3-5}
    & & $L^{(3)}$ & \pbox{\linewidth}{$98.9\%$ \\ \scriptsize{with $N_{sens} = 12$}} & \pbox{\linewidth}{$99.4\%$ \\ \scriptsize{with $N_{sens} = 3$}} \\ \hline
    \textbf{4.} & \pbox{5cm}{Trajectory accuracy in\\controlled environment} & $A^{(4)}$ & \pbox{\linewidth}{$73\%$ \\ \scriptsize{at $S = 150$ m$^2$} \\ \scriptsize{and $d_{min}\approx 20\;$cm}} &  \pbox{\linewidth}{$77\%$ \\ \scriptsize{at $S = 50$ m$^2$} \\ \scriptsize{and $d_{min}\approx 20\;$cm}} \\ \hline
    \textbf{5.} & \pbox{5cm}{Trajectory accuracy in\\real-life environment} & $A^{(5)}$ & NA & \pbox{\linewidth}{$79\%$ \\ \scriptsize{at $S_{in} = 38\;$m$^2$}}                                                
                    
  \end{tabular}
  \caption{Table with the aggregated results of each test for both
    tracking setups. The table contains a column for each test. The
    top two rows show the name of the test and the metric used to
    score the test. Underneath we have two rows For each tracking
    setup indicating how the setup scored and the difficulty of the
    test.}
  \label{tab: results}
\end{table}  }

\newpage
\section{acknowledgements}
  This work is part of the HTSM research program “HTCrowd: a
  high-tech platform for human crowd flows monitoring, modeling and
  nudging” with project number 17962, and the VENI-AES research
  program “Understanding and controlling the flow of human crowds”
  with project number 16771, both financed by the Dutch Research
  Council (NWO). The authors want to thank Dr. Antal Haans and
  Dr. Philip Ross for their effort in the establishment of the TU/e tracking setup. 
% \end{acknowledgements}

\bibliography{template}

\newpage
\appendix
\section{Counting algorithm}
\label{app: counting_algo}
In this appendix we describe the algorithm that we used to quantify
the number of pedestrians crossing a line in the TU/e setup for test
1.

\begin{figure}[H]
  \centering
  \includegraphics[width=0.495\linewidth]{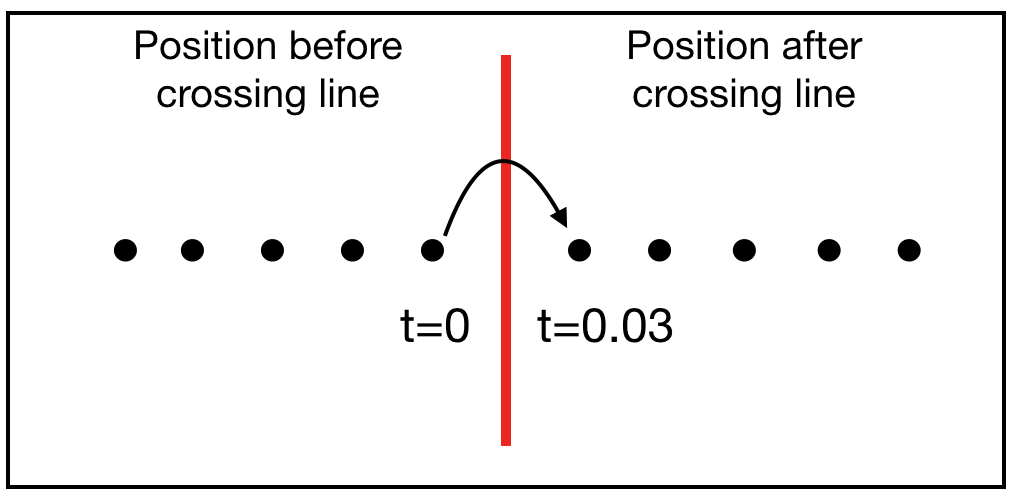}
  \includegraphics[width=0.495\linewidth]{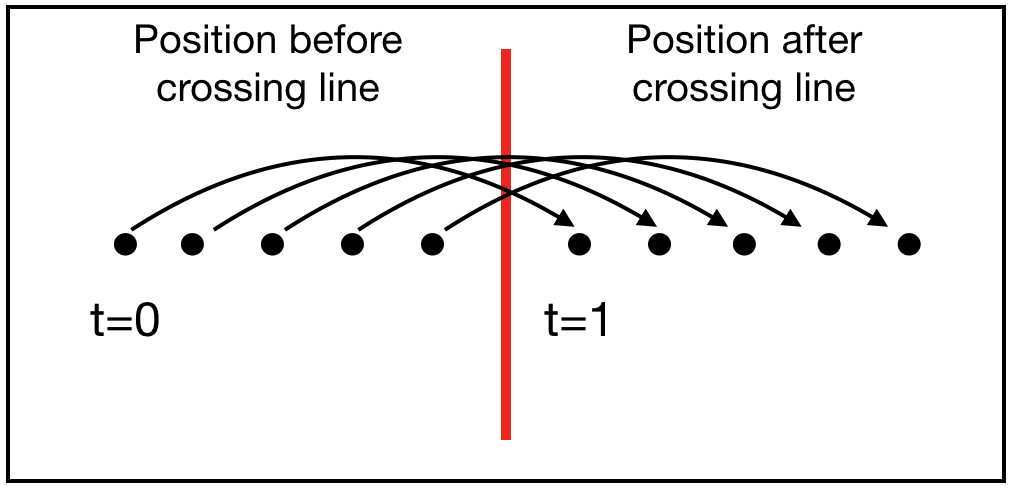}
  \caption{Conceptual sketch of algorithms detecting the crossing of a
    virtual line (in red). Black dots indicate position measurements.
    (a) Conventional counting algorithm probing for two consecutive
    measurements on either side of the (virtual) crossing line. At
    $f = 30\;$ frames per seconds the measurements are $\Delta T = 0.03\;$s
    apart. (b) Procedure to determine line crossings that we used in
    the TU/e setup for test 1. See explanation in Appendix~\ref{app:
      counting_algo}.}
    \label{fig: counting_algo}
\end{figure}

A na\"ive counting algorithm could verify whether two consecutive
measurements land on opposite sides of a (virtual) reference line (see
Fig.~\ref{fig: counting_algo}a). Such an approach is highly sensitive
to measurement noise and could yield miscounts. In our test, we
adopted a more robust approach determining a line crossing event based
on multiple samples before and after the line. Specifically, we
consider pairs of measurements along the same trajectory that are
$\Delta T = 1\;$s apart (i.e. $30$ samples). A line crossing event
was triggered when, within the $\Delta T$ time interval, the majority of
pairs were located on different sides of the line (see Fig.~\ref{fig:
  counting_algo}b).

\end{document}

%% file: exp_table/experiments-table-portrait.tex
{\renewcommand{\arraystretch}{1.6}
  \begin{longtable}{|@{}c@{}|m{4.5cm}|m{5.5cm}|} \hline
    \Large{\textbf{Features}} & \centering{\Large{\textbf{Illustration}}} &
    \Large{\textbf{Description}} \\ \hline \hline
%################################################# 1

    \begin{tabular}{m{1.3cm}|m{2cm}} 
      \textbf{Test} & Crowd-flux estimation \\ \hline
      \textbf{Metric} & $A^{(1)}$ (Eq.~\ref{eq: exp1}) \\ \hline
      \rowcolor{gray!20!white}\textbf{$\Delta T$} & 5 $\times$ 1 min \\ \hline
      \textbf{N} & 12 \\ 
    \end{tabular}  & \parbox{\linewidth}{
                     \includegraphics[trim = 0 0 0 0, clip, width =
                     .9\linewidth]{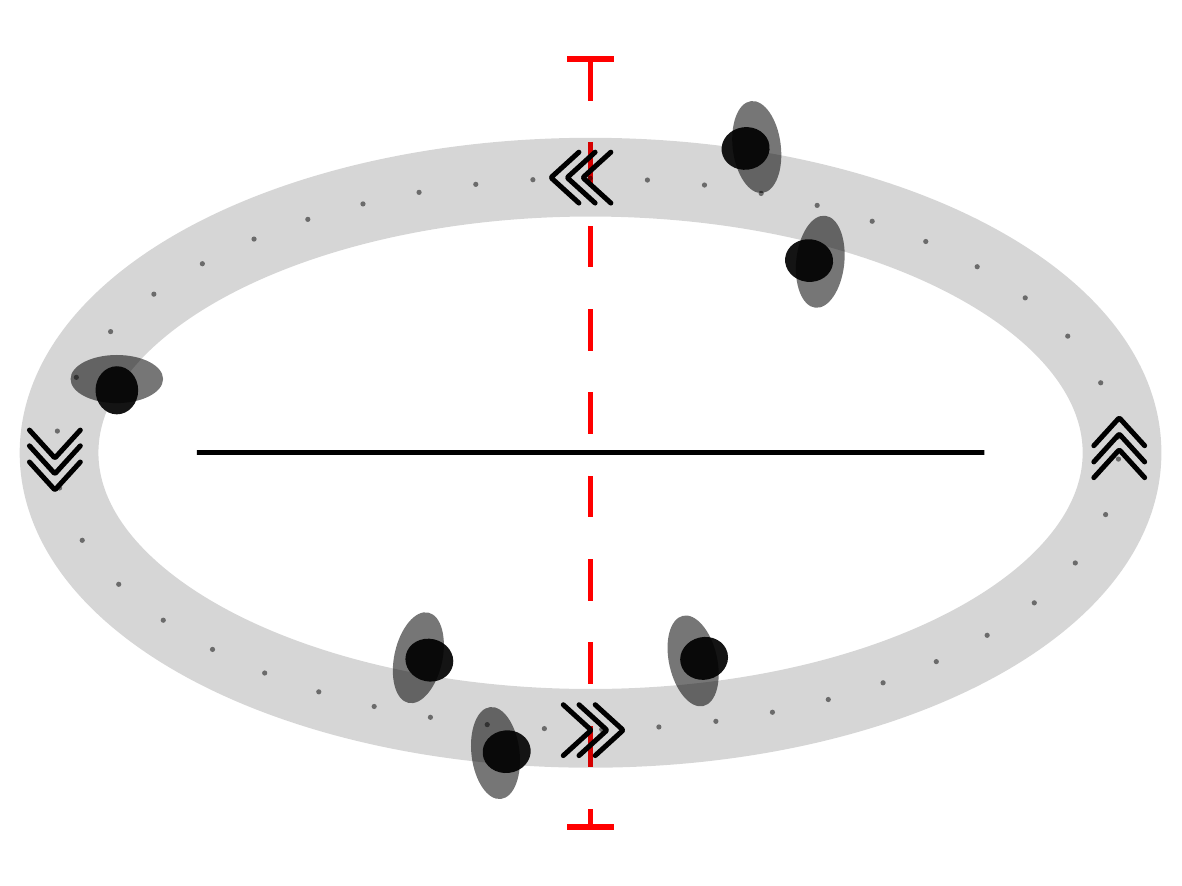}}
  &                                                     
    Participants walk in a circular path, thereby crossing a virtual
    line (in red). The test reports for every minute the error between the
    sensor estimated and the ground-truth crowd-flux across the (red)
    virtual line. \\ \hline % \hline
%################################################# 2
    
    \begin{tabular}{m{1.3cm}|m{2cm}} 
      \textbf{Test} & Density $\quad$ estimation \\ \hline
      \textbf{Metric} & $\epsilon^{(2)}$ (Eq.~\ref{eq: exp2}) \\ \hline
      \rowcolor{gray!20!white}\textbf{$\Delta T$} & $4 \times 5$ min \\ \hline
      \textbf{N} & 12 \\ 
    \end{tabular} & \parbox[c]{\linewidth}{
                    \includegraphics[trim = 0 10 0 10, clip, width=.9\linewidth]{
                      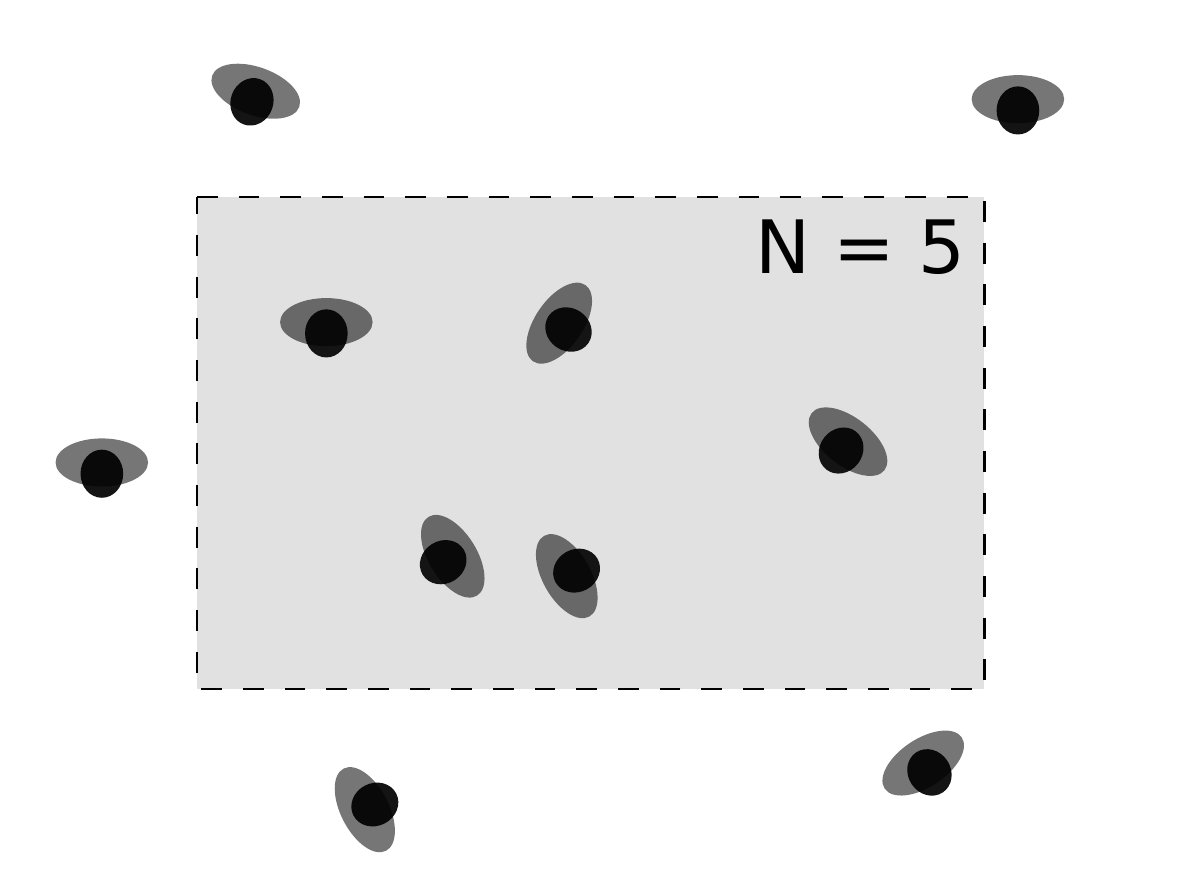}}
   &
     The number of participants inside a predefined area is kept
     constant (e.g. N = 8). The test reports the relative error between the
     estimated, $N(t)$, and ground-truth number of pedestrians, $N_{real}$, inside the area. \\ \hline% \hline 

    % ################################################# 3&4

\begin{tabular}{m{1.3cm}|m{2cm}}
  \multirow{2}{*}{\textbf{Test}} & \multirow{2}{*}{\pbox{\textwidth}{Individual\\position\\detection}} \\
   & \\ \hline
  \multirow{3}{*}{\textbf{Metrics}} & $\sigma^{(3)}$ (Eq.~\ref{eq: exp3}) \\
                                 & $D^{(3)}$ (Eq.~\ref{eq: exp3d}) \\
  & $L^{(3)}$ (Eq.~\ref{eq: exp3l}) \\ \hline
  \rowcolor{gray!20!white}$\Delta T$ & $ 2 \times 5$ min \\ \hline
  \textbf{N} & 12 \\
\end{tabular}  &
\parbox{\linewidth}{
                    \includegraphics[trim = 0 0 0 0, clip, width=.9\linewidth]{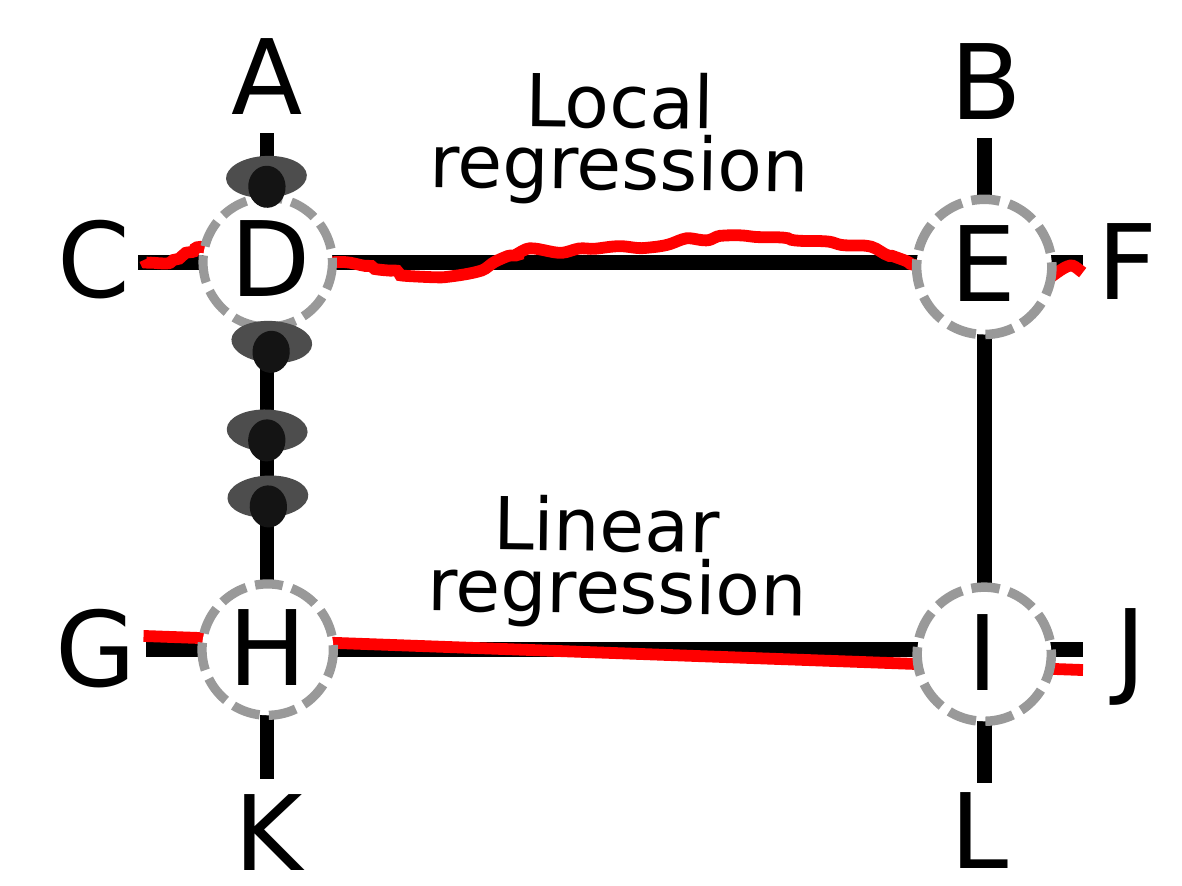}}
  & Participants walk in a row in a straight line. The test reports
    the standard deviation in the distance between the recorded data
    point and the local and linear regressions. Additionally, we report the angles and distances between the linear regressions to identify distortions in the measurement setup.\\ \hline

% ################################################# 5

\begin{tabular}{m{1.3cm}|m{2cm}}
  \textbf{Test} & Trajectory accuracy \\ \hline
  \textbf{Metric} & $A^{(4)}$ (Eq.~\ref{eq: exp4})  \\ \hline
%  \textbf{Env} & Controlled \\ \hline
  \rowcolor{gray!20!white}  \textbf{$\Delta T$} & $4 \times 1$ min \\ \hline
 % \textbf{Area} & $\approx 100 \; m^2$ \\ \hline
  \textbf{N} & 12  \\
\end{tabular}  & 
                 \parbox{\linewidth}{
  \includegraphics[trim = 0 0 0 0, clip, width=
  \linewidth]{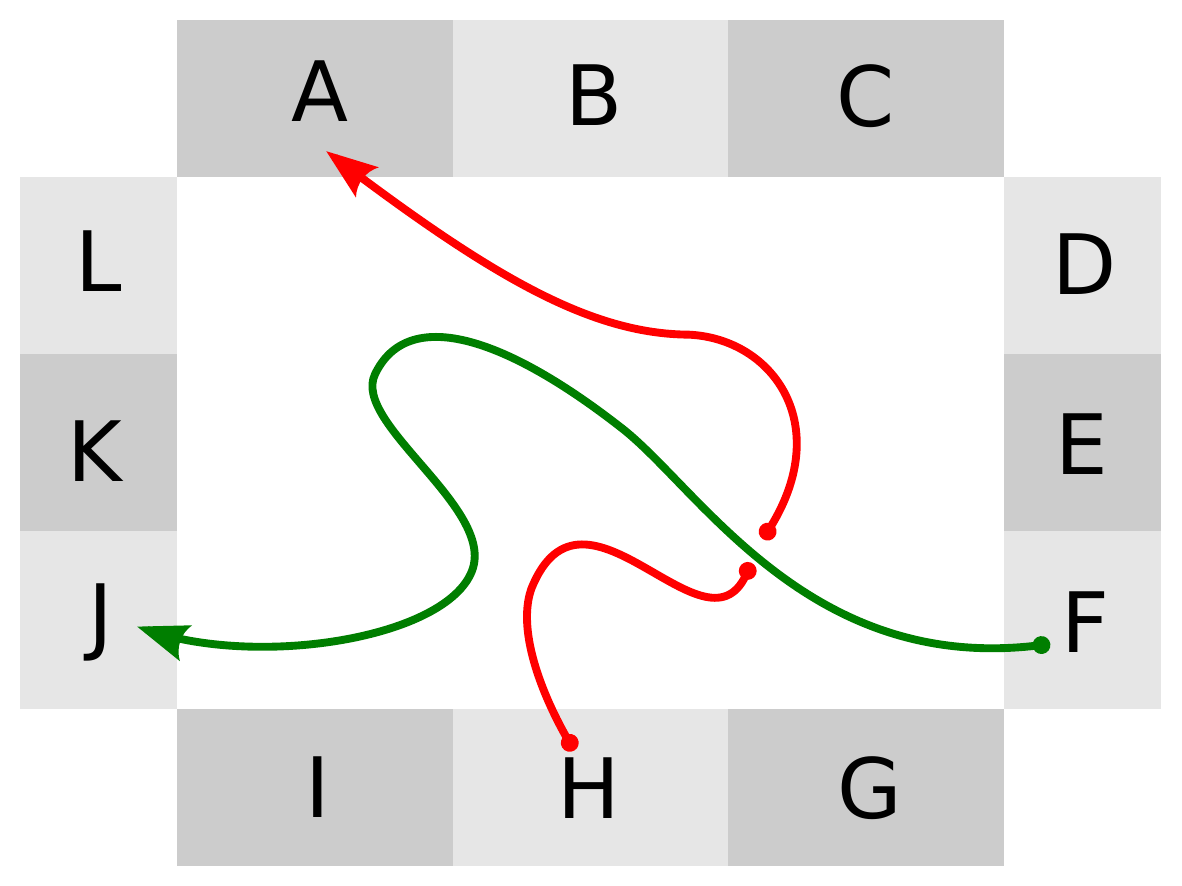}}
  & Participants walk in an irregular path from a predefined origin to
    a predefined destination, passing in close proximity from each other. The test reports the percentage of
    trajectories that is accurately tracked from origin to destination without
    interruption. \\ \hline

% ################################################# 6

\begin{tabular}{m{1.3cm}|m{2cm}}
  \textbf{Test} & Trajectory accuracy \\ \hline
  \textbf{Metric} & $A^{(5)}$ (Eq.~\ref{eq: exp5}) \\ \hline
  \textbf{Env} & Real-life \\ \hline
  \rowcolor{gray!20!white}\textbf{$\Delta T$} & $\geq$ 1 day  \\
\end{tabular}  & 
                 \parbox{\linewidth}{
                 \includegraphics[trim = 0 0 0 30, clip, width=\linewidth]{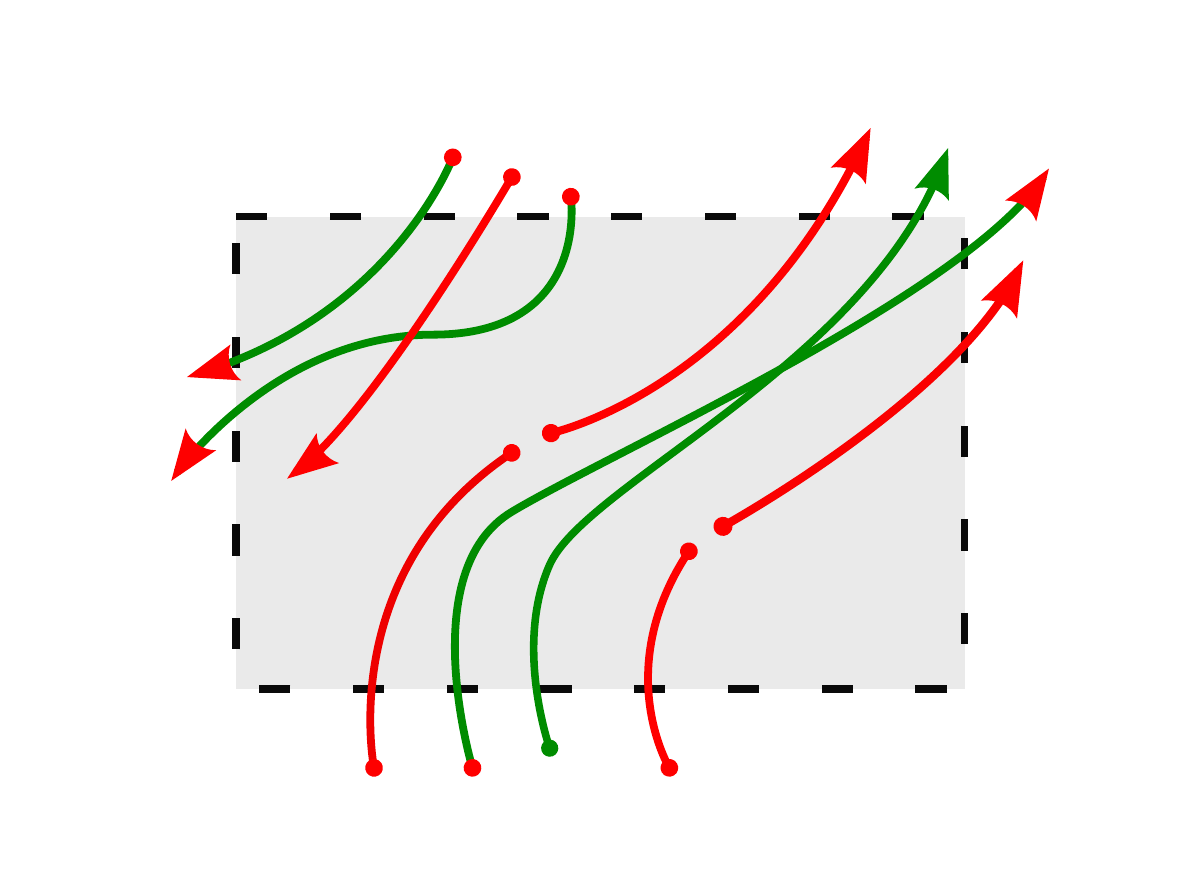}}
                 
& We define a domain, well within the sensor range, where no
trajectory should originate or terminate. The test reports in real-life conditions the percentage of
trajectories that is continuously tracked through that domain. \\ \hline
\caption{Synthetic description of the benchmark tests including key
    features and an illustration. The metric used to score the test which
    we supplemented with advised quantities for the test duration, $\Delta T$,
    and number of participants, $N$, is included. Cells with advised quantities have
    a gray background.}
\label{tab: all-tests}
\end{longtable}}

%%% Local Variables:
%%% mode: latex
%%% TeX-master: t
%%% End:

%% file: paper-arxiv.bbl
\begin{thebibliography}{10}

\bibitem{piereringaprorail}
P.~Eringa, ``Prorail: Meer en snellere treinen,'' 2020.
\newblock Accessed: 2021-07-28.

\bibitem{Daamen2003}
W.~Daamen and S.~P. Hoogendoorn, ``Experimental research of pedestrian walking
  behavior,'' {\em Transportation Research Record}, vol.~1828, pp.~20--30, 1
  2003.

\bibitem{Seyfried2005}
A.~Seyfried, B.~Steffen, W.~Klingsch, and M.~Boltes, ``{The fundamental diagram
  of pedestrian movement revisited},'' {\em Journal of Statistical Mechanics:
  Theory and Experiment}, vol.~2005, p.~10002, oct 2005.

\bibitem{Kretz2006}
T.~Kretz, A.~Gr{\"{u}}nebohm, M.~Kaufman, F.~Mazur, and M.~Schreckenberg,
  ``{Experimental study of pedestrian counterflow in a corridor},'' {\em
  Journal of Statistical Mechanics: Theory and Experiment}, vol.~2006,
  p.~P10001, oct 2006.

\bibitem{Moussad2009}
M.~Moussad, D.~Helbing, S.~Garnier, A.~Johansson, M.~Combe, and G.~Theraulaz,
  ``Experimental study of the behavioural mechanisms underlying
  self-organization in human crowds,'' {\em Proceedings of the Royal Society B:
  Biological Sciences}, vol.~276, pp.~2755--2762, 8 2009.

\bibitem{Schadschneider2011}
A.~Schadschneider, W.~Klingsch, H.~Kl{\"{u}}pfel, T.~Kretz, C.~Rogsch, and
  A.~Seyfried, ``{Evacuation Dynamics: Empirical Results, Modeling and
  Applications},'' in {\em Extreme Environmental Events}, pp.~517--550,
  Springer, New York, NY, 2011.

\bibitem{Seitz2012}
M.~J. Seitz and G.~K{\"{o}}ster, ``{Natural discretization of pedestrian
  movement in continuous space},'' {\em Physical Review E - Statistical,
  Nonlinear, and Soft Matter Physics}, vol.~86, no.~4, 2012.

\bibitem{Yamamoto2019}
H.~Yamamoto, D.~Yanagisawa, C.~Feliciani, and K.~Nishinari, ``{Body-rotation
  behavior of pedestrians for collision avoidance in passing and cross flow},''
  {\em Transportation Research Part B: Methodological}, vol.~122, pp.~486--510,
  apr 2019.

\bibitem{Corbetta2014}
A.~Corbetta, L.~Bruno, A.~Muntean, and F.~Toschi, ``High statistics
  measurements of pedestrian dynamics,'' {\em Transportation Research
  Procedia}, vol.~2, pp.~96--104, 1 2014.

\bibitem{Brscic2014}
D.~Br{\v{s}}{\v{c}}i{\'{c}}, F.~Zanlungo, and T.~Kanda, ``{Density and velocity
  patterns during one year of pedestrian tracking},'' in {\em Transportation
  Research Procedia}, vol.~2, pp.~77--86, 10 2014.

\bibitem{Zanlungo2017}
F.~Zanlungo, Z.~Y{\"{u}}cel, D.~Br{\v{s}}{\v{c}}i{\'{c}}, T.~Kanda, and
  N.~Hagita, ``{Intrinsic group behaviour: Dependence of pedestrian dyad
  dynamics on principal social and personal features},'' {\em PLoS ONE},
  vol.~12, nov 2017.

\bibitem{Corbetta2018}
A.~Corbetta, J.~A. Meeusen, C.-M. Lee, R.~Benzi, and F.~Toschi,
  ``{Physics-based modeling and data representation of pairwise interactions
  among pedestrians},'' {\em Physical Review E}, vol.~98, dec 2018.

\bibitem{Willems2020}
J.~Willems, A.~Corbetta, V.~Menkovski, and F.~Toschi, ``{Pedestrian orientation
  dynamics from high-fidelity measurements},'' {\em Scientific Reports 2020
  10:1}, vol.~10, pp.~1--10, jul 2020.

\bibitem{Pouw2020}
C.~A.~S. Pouw, F.~Toschi, F.~van Schadewijk, and A.~Corbetta, ``Monitoring
  physical distancing for crowd management: Real-time trajectory and group
  analysis,'' {\em PLOS ONE}, vol.~15, 10 2020.

\bibitem{Boltes2013}
M.~Boltes and A.~Seyfried, ``Collecting pedestrian trajectories,'' {\em
  Neurocomputing}, vol.~100, pp.~127--133, 1 2013.

\bibitem{Brscic2013}
D.~Br{\v{s}}{\v{c}}i{\'{c}}, T.~Kanda, T.~Ikeda, and T.~Miyashita, ``{Person
  tracking in large public spaces using 3-D range sensors},'' {\em IEEE
  Transactions on Human-Machine Systems}, vol.~43, pp.~522--534, nov 2013.

\bibitem{Seer2014}
S.~Seer, N.~Brändle, and C.~Ratti, ``Kinects and human kinetics: A new
  approach for studying pedestrian behavior,'' {\em Transportation Research
  Part C: Emerging Technologies}, vol.~48, pp.~212--228, 11 2014.

\bibitem{Yoshimura2014}
Y.~Yoshimura, S.~Sobolevsky, C.~Ratti, F.~Girardin, J.~P. Carrascal, J.~Blat,
  and R.~Sinatra, ``{An analysis of visitors' behavior in the louvre museum: A
  study using bluetooth data},'' {\em Environment and Planning B: Planning and
  Design}, vol.~41, no.~6, pp.~1113--1131, 2014.

\bibitem{Centorrino2021}
P.~Centorrino, A.~Corbetta, E.~Cristiani, and E.~Onofri, ``Managing crowded
  museums: Visitors flow measurement, analysis, modeling, and optimization,''
  {\em Journal of Computational Science}, vol.~53, p.~101357, 7 2021.

\bibitem{Hong2018}
H.~Hong, G.~D. De~Silva, and M.~C. Chan, ``Crowdprobe: Non-invasive crowd
  monitoring with wifi probe,'' in {\em Proceedings of the ACM Interactactive,
  Mobile, Wearable and Ubiquitous Technologies}, vol.~2, (New York, NY, USA),
  p.~23, Association for Computing Machinery, 9 2018.

\bibitem{Georgievska2019}
S.~Georgievska, P.~Rutten, J.~Amoraal, E.~Ranguelova, R.~Bakhshi, B.~L.
  de~Vries, M.~Lees, and S.~Klous, ``Detecting high indoor crowd density with
  wi-fi localization: a statistical mechanics approach,'' {\em Journal of Big
  Data}, vol.~6, 12 2019.

\bibitem{Heuvel2019}
J.~van~den Heuvel, J.~Thurau, M.~Mendelin, R.~Schakenbos, M.~van Ofwegen, and
  S.~P. Hoogendoorn, ``An application of new pedestrian tracking sensors for
  evaluating platform safety risks at swiss and dutch train stations,'' in {\em
  Traffic and Granular Flow '17} (S.~H. Hamdar, ed.), pp.~277--286, Springer
  International Publishing, 2019.

\bibitem{Thurau2019}
J.~Thurau, J.~van~den Heuvel, N.~Keusen, M.~van Ofwegen, and S.~P. Hoogendoorn,
  ``{Influence of Pedestrian Density on the Use of the Danger Zone at Platforms
  of Train Stations},'' in {\em Traffic and Granular Flow '17}, pp.~287--296,
  Springer International Publishing, 10 2019.

\bibitem{Thurau2020}
J.~Thurau and N.~Keusen, ``{Influence of obstacles on the use of the danger
  zone on railway platforms},'' {\em Collective Dynamics}, vol.~5, p.~A84, aug
  2020.

\bibitem{kinect}
M.~Coorporation.
\newblock Microsoft Coorporation, Kinect for Xbox 360, Redmond WA, USA.

\bibitem{Kroneman2020}
W.~Kroneman, A.~Corbetta, and F.~Toschi, ``Accurate pedestrian localization in
  overhead depth images via height-augmented hog,'' {\em Collective Dynamics},
  vol.~5, 3 2020.

\bibitem{Corbetta2020}
A.~Corbetta, W.~Kroneman, M.~Donners, A.~Haans, P.~Ross, M.~Trouwborst, S.~{Van
  de Wijdeven}, M.~Hultermans, D.~Sekulovski, F.~{Van der Heijden}, S.~Mentink,
  and F.~Toschi, ``{A large-scale real-life crowd steering experiment via
  arrow-like stimuli},'' {\em Collective Dynamics}, vol.~5, pp.~61--68, mar
  2020.

\bibitem{trackpy}
D.~B. Allan, T.~Caswell, N.~C. Keim, C.~M. van~der Wel, and R.~W. Verweij,
  ``soft-matter/trackpy: Trackpy v0.5.0,'' Apr. 2021.

\bibitem{Corbetta2017}
A.~Corbetta, C.-m.~M. Lee, R.~Benzi, A.~Muntean, and F.~Toschi, ``{Fluctuations
  around mean walking behaviors in diluted pedestrian flows},'' {\em Physical
  Review E}, vol.~95, mar 2017.

\bibitem{Liu2009}
X.~Liu, W.~Song, and J.~Zhang, ``{Extraction and quantitative analysis of
  microscopic evacuation characteristics based on digital image processing},''
  {\em Physica A: Statistical Mechanics and its Applications}, vol.~388,
  pp.~2717--2726, jul 2009.

\end{thebibliography}
